\documentclass[twocolumn,twocolappendix]{aastex631}

\usepackage{amsmath}
\usepackage{amssymb}
\usepackage{color}
\usepackage{graphicx}
\usepackage{listings}
\usepackage{comment}
\usepackage{hyperref}

\graphicspath{{plots/}}

\newcommand{\code}[1]{\texttt{#1}}
\newcommand{\appUIref}[1]{Figure~\ref{f:app_controls}#1}
\newcommand{\pgUIref}[1]{Figure~\ref{f:particle_controls}#1}

\newcommand{\thedocumentation}[1]{\href{http://www.alexbgurvi.ch/Firefly/docs/build/html/#1}{the documentation}}
\newcommand{\docref}[2]{\href{http://www.alexbgurvi.ch/Firefly/docs/build/html/#1}{\code{#2}}}

\def\threejs/{\href{https://threejs.org/}{\code{three.js}}}
\def\dthreejs/{\href{https://d3js.org/}{\code{d3.js}}}
\def\kaitaiio/{\href{https://kaitai.io}{\code{kaitai.io}}}
\def\matplotlib/{\href{https://matplotlib.org/stable/index.html}{Matplotlib}} 
\def\flask/{\href{https://flask.palletsprojects.com/en/2.1.x/}{\code{Flask}}}
\def\website/{\href{https://alexbgurvi.ch/Firefly}{alexbgurvi.ch/Firefly}}
\def\particlegroup/{\docref{data\_reader/reader.html\#the-particlegroup-class}{firefly.ParticleGroup}}
\def\json/{\code{.JSON}}
\def\ffly/{\code{.ffly}}

\begin{document}

\title{Firefly: a browser-based interactive 3D data visualization tool for millions of data points}

%\newcommand{\section}[1]{\vspace{1em}\noindent\textit{#1}\\ \indent}

%% authors ------
\correspondingauthor{Alexander B. Gurvich}
\shorttitle{Firefly}
\shortauthors{Gurvich and Geller}

\author[0000-0002-0786-7307]{Alexander B. Gurvich}
\affiliation{Department of Physics \& Astronomy and CIERA, Northwestern University, 1800 Sherman Ave, Evanston, IL 60201, USA}
\email{agurvich@u.northwestern.edu}

\author{Aaron M. Geller}
\affiliation{Department of Physics \& Astronomy and CIERA, Northwestern University, 1800 Sherman Ave, Evanston, IL 60201, USA}
\affiliation{Research Computing Services, Northwestern IT, Evanston, IL 60201, USA}

\begin{abstract}
    
    We present Firefly, a new browser-based interactive tool for visualizing 3D particle data sets.  
    On a typical personal computer, Firefly can simultaneously render and enable real-time interactions with $\gtrsim$10 million particles, and can interactively explore datasets with billions of particles using the included custom-built octree render engine.
    Once created, viewing a Firefly visualization requires no installation and is immediately usable in most modern internet browsers simply by visiting a URL.
    As a result, a Firefly visualization works out-of-the-box on most devices including smartphones and tablets.
    Firefly is primarily developed for researchers to explore their own data, but can also be useful to communicate results to researchers/collaborators and as an effective public outreach tool.
    Every element of the user interface can be customized and disabled, enabling easy adaptation of the same visualization for different audiences with little additional effort. 
    Creating a new Firefly visualization is simple with the provided Python data pre-processor (PDPP) that translates input data to a Firefly-compatible format and provides helpful methods for hosting instances of Firefly both locally and on the internet.
    In addition to visualizing the positions of particles, users can visualize vector fields (e.g., velocities) and also filter and color points by scalar fields.
    We share three examples of Firefly applied to astronomical datasets: 1) the FIRE cosmological zoom-in simulations, 2) the SDSS galaxy catalog, and 3) Gaia DR3.
    A gallery of additional interactive demos is available at \href{https://alexbgurvi.ch/Firefly}{alexbgurvi.ch/Firefly}.
\end{abstract}


\section{Introduction}

    As datasets of all kinds become larger and more complex, exploring and extracting information from them has become commensurately more difficult.
    In particular, with advancements in modern computer processors and data collecting instruments, the size of astronomical datasets has grown by orders of magnitude in the last two decades (e.g. SDSS: \citealt{York2000}, \citealt{Blanton2017}; Gaia: \citealt{GaiaCollaboration2016}, \citealt{Babusiaux2022}; anticipated data output of V. Rubin Observatory \citealt{Ivezic2019})
    The scale and complexity of these massive datasets makes it difficult to grasp their full scientific content through traditional analysis and statistical methods alone.

    Hydrodynamic cosmological simulations are an important example of a complex dataset in astronomy that benefits from 3D visualization.
    They are extremely rich in information, and the larger the volume (or the higher the resolution) the more structures that can be easily missed by standard pipeline analysis tools.
    Their output typically consists of 3D spatial coordinates annotated with various scalar quantities for each resolution element. 
    However, as the number of resolution elements in these simulations increases, interactive visualization becomes increasingly more difficult. 
    
    A common visualization workflow for these kinds of simulations is to produce a series of 2D projections or slices of the 3D dataset \citep[using e.g. \code{FIRE Studio},][]{Gurvich2022a}.
    These 2D images are often stitched together from a rotating perspective into an animation, to give a sense of three-dimensionality.
    However, constructing a publication-quality static image (or movie) often requires \textit{a priori} knowledge of the most important portion and features of the data to view.
    Additionally, relevant science questions often requires the data to be filtered by one quantity (e.g., only showing very dense material) and then colored by another quantity (e.g., temperature).
    It's often the case that the ``best'' parameters for how to filter and/or colormap the data are not known, and the best 2D images can only be obtained after extensively exploring the parameter space.
    However, producing static images (and/or movies) for all the different combinations of filter and colormap parameters can be extremely time consuming and inefficient.
    Rendering individual frames for these movies can also take an enormous amount of computational effort, and changing an aspect of the visualization often requires re-rendering from scratch, making it prohibitive to iteratively refine the visualization at full scale.
    This leaves little room for scientific discovery.
    A more efficient visualization workflow is to \textit{interactively} examine the dataset in real time and freely explore the different quantities. 
    
    Interactive visualization enables users to interrogate a dataset in real time, e.g., through manipulating the camera location, filtering out portions of data, coloring portions of data differently, etc.  
    Moreover, interactive visualization is a powerful tool for building intuition about a dataset and enabling serendipitous scientific discovery, which can be particularly important for large and complex datasets.
    However, effective interactive visualization is technically challenging, as it requires the ability to render dozens of frames per second while also reacting to user input, and therefore the field has struggled to keep up with the pace at which data has ballooned.
    In addition to the technical hurdles, it is also challenging to design visualization software that is both accessible to creators and end-users.  In this paper we present a new general-purpose web-based interactive visualization tool, Firefly\footnote{This package should not be confused with the serendipitously named web-based visualization software firefly, from Caltech-IPAC \href{https://github.com/Caltech-IPAC/firefly}{github.com/Caltech-IPAC/firefly}), a general tool for retrieving and viewing astronomy FITS files.}, that addresses many of these visualization challenges and draws from the successes of previous visualization software.

    One field that has contributed greatly to the 3D interactive visualization landscape is photogrammetry, which visualizes point-based data taken from Laser Imaging, Detection, and Ranging (LIDAR) equipped drones. 
    Modern photogrammetric datasets can contain as many as 43 trillion data points \footnote{See for example the public dataset of LIDAR readings across the US provided by the US Geological Service at: \href{https://registry.opendata.aws/usgs-lidar/}{registry.opendata.aws/usgs-lidar} which can be interactively explored at: \href{https://usgs.entwine.io/}{usgs.entwine.io}}.
    A popular approach for visualizing these datasets is to use highly optimized point-based rendering methods \citep[e.g.][]{Martinez-Rubi2015,Schutz2016}. 
    These special purpose softwares are well optimized to handle hundreds of billions of data points at a time to produce interactive topographical maps\footnote{See for example \href{https://usgs.entwine.io/data/view.html?r=\%22https://s3-us-west-2.amazonaws.com/usgs-lidar-public/USGS_LPC_IL_4County_Cook_2017_LAS_2019\%22}{this interactive visualization} of Cook County Illinois}.
    We borrow and adapt some of their successful techniques (e.g., progressive rendering of massive particle data sets, and browser-based viewers) in Firefly.
    
    Unlike many other fields in science, Astronomy has a highly heterogenous audience and data structure. Astronomers span from observers to theorists to computer modelers to instrumentation engineers (and any combination of these).  Astronomy also has a historical ability to engage wide swaths of the public.  Furthermore, the wide range of data taken across the electromagnetic spectrum (from different telescopes) and differences in modeling/simulation codes produces a similarly wide range in data formats and requirements.
    
    Thus, many astronomy tools have been developed to visualize specific data, each filling it's own niche.  For example, much of the current software for visualizing Gaia data is purpose built in order to handle the data volume and to interface directly with the Gaia archive (e.g.,
    the Gaia archive visualization service \citealt{Moitinho2017} and \code{VAEX} \citealt{Breddels2018}).
    
    There are also many existing tools for general-purpose (interactive) visualization of different astronomical datasets. A non-exhaustive list includes \code{Partiview} \citep{Levy2003}, \code{yt} \citep{Turk2011},  \code{ParaView} \citep{Ayachit2015}, \code{pysphviewer}  \citep{Benitez-Llambay2015}, \code{IGM-Vis} \citep{Burchett2019}, and \code{Polyphorm} \citep{Elek2020}.
    Each software fills a niche in its community and, as a result, has its own interface, data structures, and limitations.
    One approach to bridge the gap between communities is \code{Glue} \citep{glueviz}, which creates a unified interface and data input framework for multiple specialized visualization and analysis softwares using ``plugins''.  Python's  \matplotlib/ \citep[][]{Hunter2007} is another very popular tool in the astronomy community for visualizing modest sized data sets (e.g., thousands of points). 
    %While purpose built web-based point cloud visualization software like for LIDAR is available general purpose web-baesd point cloud visualization is limited to products like \href{https://scenemark.com/}{\code{Scenemark}}.

    In creating Firefly we have attempted to include many of the successful features of previous tools while also addressing needs that we were unable to find in existing tools. Our aim is to create software that:
    \begin{itemize}
        \item supports any array data
        \item is easy to ingest data into using an interface that many astronomers are already familiar with (Python)
        \item can interactively render up to tens of millions of points in its default mode, and billions of points by progressively loading data as necessary
        \item produces visualizations accessible by url which can be viewed without any installation on a wide array of devices including smartphones, tablets, and computers
        \item has a customizable interface to create visualizations for different audiences so that the same data can be explored by a researcher, presented to peers, and presented to the public with minimal additional effort
    \end{itemize}
    
    Firefly enables interrogation of large data sets ($\gtrsim 10^6-10^9$ points) in real time from within a web browser without any installation and
    allows for customization of aspects of the visualization and elements of the user interface through shareable app-state configuration files to tailor visualizations for any audience.
    Firefly is configured to visualize \textit{any} three-dimensional dataset (though it will likely produce the best results if those dimensions are spatial x,y,z). 
    Firefly also has specific tools for visualizing particle velocities (a common attribute of many observed and simulated astronomical data sets) or any other vector field.
    Users can also specify scalar fields to color/filter/scale the particles. 
    In this sense, Firefly is flexible enough to be applied in any context in which there are three dimensions that one would like to visualize in a web browser.  

    In the following sections, we describe how to use Firefly and provide an in-depth description of the code.
    In \S \ref{s:general_use_cases} we use illustrative example use-cases to highlight the key features of Firefly. 
    In \S \ref{s:astro_datasets} we showcase example Firefly visualizations of astronomical datasets. 
    In \S \ref{s:performance_tests}, we evaluate the performance of Firefly for different sized datasets.
    Finally, in \S \ref{s:conclusion} we summarize and conclude. 

    The appendices contain detailed information about the implementation and usage of many of Firefly's features.
    In Appendix \ref{s:webapp_features} we describe how to use the various elements of the user interface in a Firefly web application.
    In Appendix \ref{s:python_pdpp} we describe how to use the provided Python data pre-processor (PDPP) to create new Firefly visualizations, generate settings files, and host instances of Firefly both locally and on the internet. 
    In Appendix \ref{s:under_the_hood} we describe the inner workings of the web application. 
    In Appendix \ref{s:octree} we describe the (optional) progressive rendering scheme for visualizing %subsets of 
    datasets which have been pre-formatted as an octree (allowing Firefly to scale to datasets of ${\gtrsim}10^9$ particles).

\section{Use cases and key features}\label{s:general_use_cases}
    Firefly is a flexible application with many features that can be enabled/disabled using the same dataset without fundamentally altering the code.
    In this section we provide some example use cases in order to highlight  key features of Firefly that can be used to create a tailored experience.
    We categorize the use cases based on the intended audience and/or purpose: interactively exploring ones' own data (\S\ref{ss:research_tool}), sharing results with other scientists and collaborators (\S\ref{ss:sharing_results}), and public outreach and broader science communication (\S\ref{ss:public_outreach}).
    
    Many of the features described below require a \flask/-enabled Firefly server, which allows communication between a live Python interpreter and a Javscript interpreter using websockets \citep[for details, see][]{Grinberg2018}.
    A \flask/-enabled Firefly server can be launched locally using the \code{firefly} command, which is made available when Firefly is \code{pip} installed, with the \code{--method=flask} command-line argument. Additional information about installing and using Firefly with Flask is available in Appendix~\ref{s:flask_backend}.
    
    Lastly, we note that the features we highlight in the sections below can easily be mixed and matched to create a compelling Firefly experience for any audience.
    Additional features and information are available in \thedocumentation/.
    To help guide the reader to the text they may be most interested in, in each of the following subsections we first provide a bulleted list of features to be discussed and then use \textbf{bold} font within the subsequent narrative to point the reader to where these specific features are mentioned.
    
    \subsection{Research tool for data exploration\label{ss:research_tool}}
        In this section we discuss the following features: 
        \begin{itemize}
            \item the Python data pre-processor (PDPP)
            \item coloring/filtering by an attribute
            \item plotting the velocity vector field
            \item moving the rotation anchor
            \item extrapolating position along velocity vectors
            \item embedding Firefly in a jupyter notebook
            \item accessing data on a remote server and streaming a scene rendered on another computer
        \end{itemize}
    
        \noindent Firefly is a powerful and versatile tool for researchers to interactively explore their own data in real time.  
        The first step in creating any Firefly visualization is to process the dataset using Firefly's built-in \textbf{Python data pre-processor} (\textbf{PDPP}; see Appendix~\ref{s:python_pdpp}).
        The Python data pre-processor can interpret a range of common data file formats (e.g. csv and hdf5; additional file formats can also be added by modifying the source code). 
        Typical data will consist of discrete particles that each have x,y,z coordinates, and any number of additional attributes (e.g., velocity, temperature, density, etc.).
        The most straight-forward procedure is to use Firefly's PDPP to save a copy of the pre-processed data to disk and then launch Firefly within a web browser (see Appendix~\ref{s:under_the_hood}).
        Firefly will then read in and visualize the pre-processed data and allow the researcher to interactively e.g., \textbf{color and/or filter data based on an attribute}.  
        
        As a concrete example, a researcher who is interested in the 3D morphology of high-temperature regions in their particle data set could filter out all particles below their temperature threshold.
        Then the researcher could toggle the camera mode from the default ``trackball controls'' (which rotates the scene) to ``fly controls''  in order to move around the data volume.
        By flying the camera into one of these high-temperature regions and switching back into trackball controls they can \textbf{reposition the rotation anchor} at the new location to visually inspect the morphology of each region and note their locations for potential future detailed follow-up investigations.
        Perhaps the researcher is also interested in the magnitudes and directions of the velocities in these high-temperature regions; in that case, they can color the points by velocity and \textbf{plot velocity vectors} rather than individual scatter points (e.g., see Figure~\ref{f:full_screengrab} (a) and (b)).
        The researcher could also select to (linearly) \textbf{extrapolate the particle positions along the direction of the velocity vectors} in a loop of a given duration to give an intuitive sense of how the material will flow over time by enabling the ``animate velocities'' checkbox.
        
        Firefly can also run while \textbf{embedded within a Jupyter notebook} (Appendix~\ref{ss:jupyter}).  
        In this mode, it is often most effective to use a \flask/-enabled server to launch Firefly, e.g., using the \code{firefly --method=flask} command from the terminal.
        However the Firefly server is launched, a Firefly visualization can then be embedded as an IFrame in the Jupyter notebook:
\begin{lstlisting}[language=python]
from IPython.display import IFrame
IFrame(`http://localhost:<port>/')
\end{lstlisting}
        Then, the researcher can work directly within the Jupyter notebook to manipulate their data, process it for Firefly, and send (POST) the data directly to their Firefly server (a task that is conveniently wrapped by the \docref{reference/api/classes/firefly.data\_reader.Reader.html\#firefly.data\_reader.Reader.sendDataViaFlask}{Reader.sendDataViaFlask} method of the Python PDPP).
        As a result, the researcher can load their data and interactively explore it in Firefly as easily as they would make static visualizations or scientific plots with the same footprint on disk (i.e,. without having to double the size of the dataset by copying it on disk in Firefly compatible format). 
    
        If the researcher's local machine does not have sufficient resources, they can connect to a high-performance computer (HPC) where they process the data and launch a Firefly server.  
        By port forwarding, the researcher could view and manipulate Firefly on their local computer while accessing the Firefly server on the HPC.  
        In this scenario, the researcher's web browser on their local machine will \textbf{read in the data stored on the HPC} without having to manually copy data over to the local machine. 
        Alternatively, the researcher could launch a web browser directly on the HPC and only ``\textbf{stream}" the visual content to their local computer (using the \code{<ip-address>:<port>/stream} entry point, which is only available when Firefly is launched with a \flask/-enabled server; note that firewalls on some centrally managed HPCs might prohibit this streaming method). 
        Here the local computer does not load in any data other than the rasterized frame-by-frame images from the HPC.   
        The quality of the rasterized image is automatically degraded in order to preserve a minimum FPS (which can be configured using the \code{--framerate=<fps>} command line argument when Firefly is launched).  
        Detailed instructions on this procedure is available in \thedocumentation/. 
    
    \subsection{Collaborative tool for sharing results with peers\label{ss:sharing_results}}
        In this section we discuss the following features:  
        \begin{itemize}
            \item mirroring a view between two computers 
            \item preset startup settings files for specific views
            \item exporting standalone Firefly instances
        \end{itemize}
        \noindent In this category of use we envision the following scenarios: 1) a researcher communicating with a collaborator live via videoconferencing software and 2) a researcher sharing the results of a recent paper on a stand-alone website.
        
        In scenario 1), a researcher can set up Firefly using the \code{--method=flask} command line argument as above. 
        At this point the researcher could simply share their screen via Zoom in order to illustrate their point using Firefly.
        However, this can often introduce latency which can be frustrating and, in the worst cases, actively confusing if the framerate is exceptionally low.
        Instead, if the researcher sets up their router to allow incoming connections through the port Firefly is served on, this would allow their collaborator to access their Firefly session by visiting the corresponding \code{<ip-address>:<port>/viewer} entry point;
        The view from the researcher's Firefly session would then be \textbf{mirrored} and rendered on their collaborator's computer.

        In scenario 2), a researcher desires to create a Firefly visualization that they can provide as supplemental data to a publication that readers can freely access asynchronously.
        To do so, the researcher can first create their Firefly visualization locally. 
        They can optionally use Firefly to generate a \textbf{start-up settings file} that will initialize the view at start-up to an exact position and with applied filters and colormap.
        This process can be repeated with different datasets, e.g., separated in time, in order to create a multi-dataset visualization. 
        We provide easy-to-use functions in Firefly's PDPP that can \textbf{export standalone Firefly instances}. 
        To export a Firefly instance the PDPP creates a new directory containing only the necessary Firefly source files and data that is ready to be uploaded to a personal web server, or to copy this local Firefly visualization to GitHub Pages (a free web-hosting service provided by GitHub).
        
        Once the visualization is hosted online, the researcher can include the url in their publication and on their own personal website to direct readers to the interactive visualization where they can learn more about the research in an engaging and unique way. 

    \subsection{Public outreach tool for science communication\label{ss:public_outreach}}
        In this section we discuss the following features:  
        \begin{itemize}
            \item splitting the view and user interface into separate windows
            \item enabling/disabling elements of the user interface
            \item pre-defining a camera path
            \item VR compatability
        \end{itemize}
    
        \noindent In this category, we envision two scenarios: 1) a researcher speaking in front of an audience and 2) a museum visitor exploring data within a Firefly visualization as part of an exhibit. 

        In scenario 1) the researcher launches Firefly with a \flask/-enabled server using the \code{--method=flask} command line argument. 
        This allows the researcher to \textbf{split the view} into two different browser windows; one with only the visualization imagery, and another with the user-interface controls.
        The researcher can access the \code{localhost:<port>/viewer} entry point on the display visible to their audience and the \code{localhost:<port>/gui} entry point on a hand-held device.
        In doing so, the researcher is able to interactively control the view from Firefly without revealing the controls interface to the audience, making for a cleaner, simpler, and more engaging viewing experience. 
        
        In scenario 2) the researcher could create a Firefly visualization and \textbf{restrict aspects of the user interface} using the settings file in order to simplify the experience and reduce the amount of ``visual clutter'' in the user-interface in order to make it easier for an inexperienced user to navigate.
        The researcher can also, optionally, \textbf{define a camera path} that viewers will automatically be flown along.
        This allows the researcher to highlight specific aspects of the visualization while still allowing the viewer to interactively change aspects of the visualization (like the colormap and filters).
        If the museum has a \textbf{VR headset} available, the \code{localhost:<port>/VR} entry point will enable basic camera controls using the orientation of a VR headset so that visitors can put on the VR headset and immerse themselves in the data.
            
\section{Example Firefly visualizations of astronomical datasets}
\begin{figure*}
    \centering
    \includegraphics[width=\linewidth]{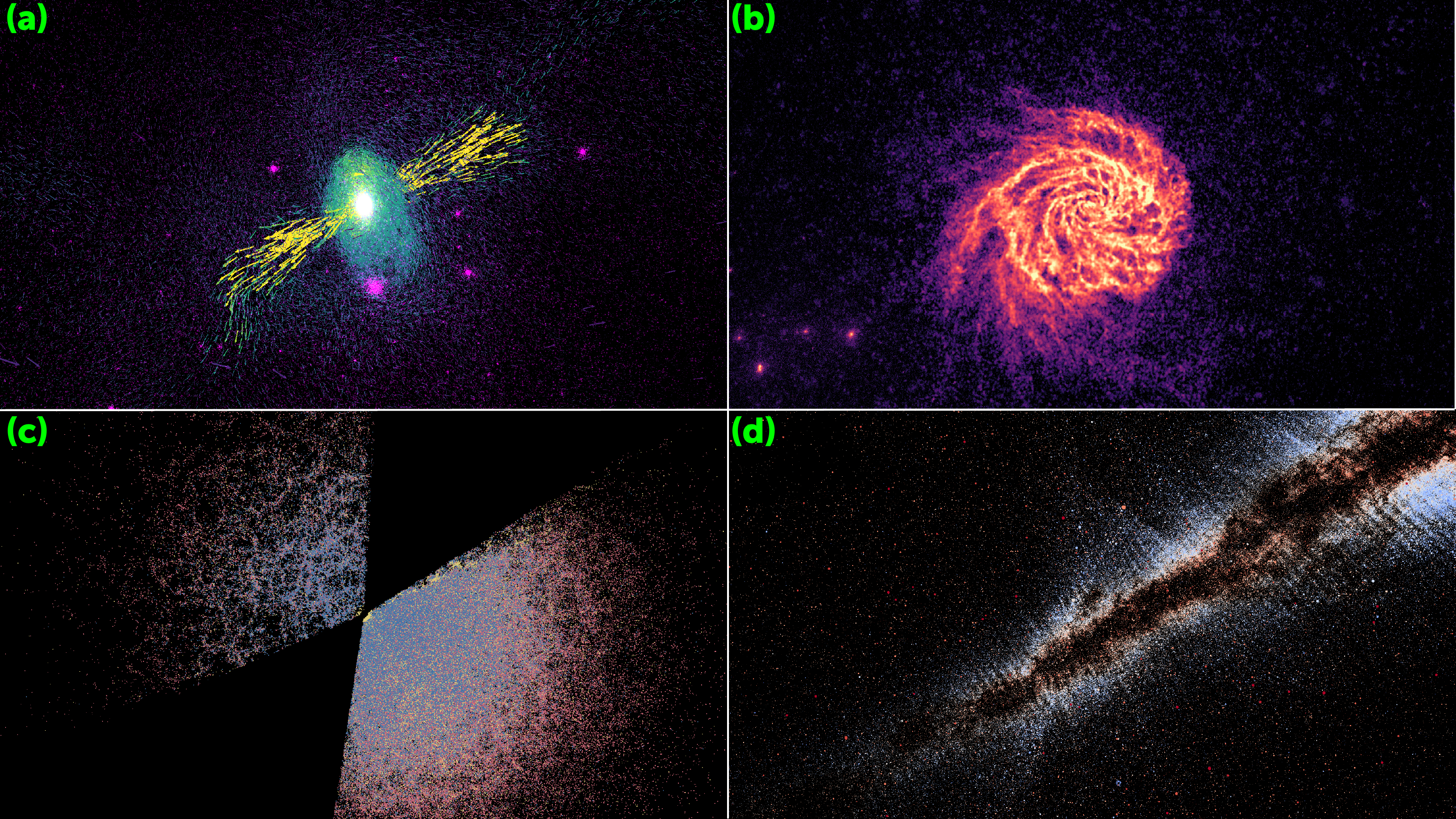}
    \caption{\label{f:full_screengrab}
    Screen captures of the Firefly web application in different modes visualizing the example datasets presented in \S \ref{s:astro_datasets}.
    \textit{(a):} 
    An example dataset from the FIRE simulations.
    In this example, we highlight the existence of a fast moving biconical outflow of hot gas by colormapping by velocity, scaling the particle sizes by temperature, and plotting velocity vectors.
    \textit{(b):} 
    A second, higher resolution, example dataset from the FIRE simulations.
    In this example, we highlight the floculent spiral structure of the galactic disk using the column density projection mode.
    \textit{(c):} The SDSS galaxy catalog, split into categories by Galaxy Zoo classifications.
    In this example, we contrast the spatial distributions of different morphologically classified galaxies by coloring each particle group (split by classification) a fixed color (red: elliptical, blue: disk, yellow: uncertain).
    \textit{(d):} A view from inside Gaia DR3, where points are individual stars colored by the difference in their measured blue band (bp) and red band (rp) magnitudes available.
    The galactic plane is clearly visible in the dirth of detected stars and the bp-rp of stars at its edges is indicative of dust reddening. 
    These visualizations are available to explore interactively at \href{https://www.alexbgurvi.ch/Firefly}{alexbgurvi.ch/Firefly}.
    }
\end{figure*}
\label{s:astro_datasets}
    To demonstrate the flexibility of Firefly, we provide three concrete examples of Firefly applied to different astronomical datasets and the Jupyter notebooks we used to create them.
    \begin{enumerate}
        \item a z=0 snapshot from the FIRE-2 simulations\footnote{See the FIRE project web site: http://fire.northwestern.edu.} \citep{Hopkins2018}
        \item a high(er) resolution counterpart snapshot from the FIRE-2 simulation public data release \citep{Wetzel2022} 
        \item the SDSS galaxy catalog from DR17 \citep{Meert2017} with morphological classification from Galaxy Zoo \citep{Lintott2008,Lintott2011}
        \item nearly 1.5$\times10^9$ stars from  Gaia DR3 \citep{GaiaCollaboration2016,Babusiaux2022} 
    \end{enumerate}
    These and other examples can be found in the gallery at \website/. 
    \subsection{FIRE simulations} \label{s:fire_example}
        For these examples, we use snapshot 600 (corresponding to redshift $z=0$) of the \code{m12b\_res57000} and \code{m12b\_res7100} simulations \citep[described in][]{Garrison-Kimmel2019}. 
        The first is a low(er) resolution counterpart of the second, which is from the FIRE-2 public data release \citep{Wetzel2022}.
        The Jupyter notebooks we use to generate these instances of Firefly are available at \href{https://github.com/agurvich/FIRE_lowres/tree/main/ntbks}{github.com/agurvich/FIRE\_lowres} and \href{https://github.com/agurvich/FIRE_hires/tree/main/ntbks}{github.com/agurvich/FIRE\_hires}.
        The high resolution simulation, \code{m12b\_res7100} , can be obtained from the \href{https://flathub.flatironinstitute.org/}{FlatHUB}, an online data repository provided by the Flatiron Institute.
        The top panels of Figure \ref{f:full_screengrab} show screenshots of these datasets (available to explore interactively at \href{https://alexbgurvi.ch/FIRE\_lowres}{alexbgurvi.ch/FIRE\_lowres} and \href{https://alexbgurvi.ch/FIRE\_hires}{alexbgurvi.ch/FIRE\_hires}).
        
        Both datasets contains 4 particle types: gas, stars, high resolution dark matter (HRDM), and low resolution dark matter (LRDM). 
        Each particle type has coordinates, velocities, and masses which we use to calculate the center of mass of the simulation box and the velocity of the center of mass. 
        For each particle type we additionally compute the distance to the center of mass (\code{GCRadius}; ``galactocentric radius'') and the magnitude of the velocity (\code{Velocity}) as additional ``derived'' scalar fields in order to track them in Firefly for filtering and colormapping.
        In addition to these two scalar fields, we also track the ($\log_{10}$ of the) \code{Temperature} and \code{Density} for the gas particles and the stellar age (\code{AgeGyr}) for the star particles.
        
        The low resolution dataset contains 6.2 million gas particles, 3.3 million star particles, 9.3 million HRDM particles and 3 million LRDM particles. 
        The high resolution dataset contains 58 million gas particles, 17 million star  particles, 70 million HRDM particles and 7 million HRDM particles.
        However, given that not all the particles are necessary to create an informative interactive visualization and that computational resources may be limited for some users, we reduce the HRDM and LRDM by a factor of 10${\times}$ (50${\times}$) for the low (high) resolution dataset.
        Additionally we apply 7${\times}$ (10${\times}$) reduction factors to the gas (star) particles for the high resolution dataset in order to fit within browser memory restrictions (in \S\ref{s:gaia_example} we explore an alternate approach to circumventing this browser memory restriction for an extremely large, ${\gtrsim}10^9$ particle, dataset). 
        Table~\ref{t:FIRE_dataset} summarizes the number of particles and scalar fields included in these Firefly visualizations, including the decimation factors where applicable. 
        
        We begin with the low resolution m12b\_res57000 example.
        To avoid obscuring the galaxy at the center of the box we initialize the HRDM \code{GCRadius} filter to hide all particles within 300 kpc; if we didn't do this then the view near the galaxy would be obscured by a volume filling screen of dark matter particles.
        We use the PDPP to apply this filter at startup, rather than doing it interactively at startup for two reasons: 1) it's more user friendly and 2) it avoids performance issues if you have a bunch of particles overlapping in the same pixel (see \S \ref{s:performance_tests} for a detailed breakdown of how this scales).
        We also colormap the gas particles according to the magnitude of their velocities and enable velocity vectors. By default the vector sizes (i.e., both their lengths and widths) are also set by the magnitude of their velocities.
        This approach of using color and size to doubly encode information is a best practice in creating accessible visualization.
        Here we instead choose to scale the vector sizes according to the $\log_{10}$ of their temperature.
        The ability to scale sizes by another quantity enables us to take this double-encoding approach for any quantity, not just velocities.
        With these settings, the visualization (see Figure~\ref{f:full_screengrab}a) highlights a fast moving biconical outflow (yellow) driven by stellar feedback from the center of the main galaxy surrounded by a swirling inflow of slower moving and cooler gas (purple) both above and below the disk plane (blue).
        Interestingly, the trajectory of the biconical outflow is clearly perturbed by the presence of the inflowing material as the jet bends noticably before continuing on to large radii (${\sim}100$s of kpc). 
        The presence of the biconical outflow (and its deflection) in this dataset went unnoticed until we produced this visualization, demonstrating the utility of 3D interactive visualization for exploration.

        For the second example (see Figure~\ref{f:full_screengrab}b), we show the high resolution counterpart m12b\_res7100. 
        Here we use the column density project mode, which emphasizes the flocculent (roughly ``fluffy'' and discontinuous) spiral structure of the galactic disk. 
        The column density projection mode counts the number of particles that overlap each pixel and applies a colormap to this per-pixel number density.
        This is critical for analyzing the high resolution dataset; otherwise the high density of particles quickly saturates the screen, making it difficult to identify structure.  By accumulating the data in each pixel and then assigning a color based on density, we produce a more meaningful visualization that can be more easily interpretted scientifically. 
        In this case we assign a meaningful color to the pixel  by colormapping the ($\log_{10}$ of the) projected number of particles in each pixel, which is proportional to the total mass along the line of sight, using the \code{magma} colormap from \matplotlib/.
        Using the column density mode, and in particular the $\log_{10}$ normalization, helps identify the surface density contrasts, which are typically $2-3$ orders of magnitude, which define the spiral arms of the galaxy.
        Without projecting and $\log_{10}$ normalizing the disk would appear uniform across its face due to the large number of particles saturating the pixel values and the spiral structure would not be visible.
        
        This high(er) resolution version uses the same initial starting conditions as the lower resolution simulation shown in panel (a), however interestingly, a biconical outflow does not appear in this higher resolution version (when viewed in column density or using the previous visualization settings used for the low resolution dataset). 
        Contrasting between these two different resolution simulations could help us understand how such outflows are launched and in what kinds of galaxies.
        Concretely, even if one was to explain this difference as a ``resolution artifact'' then, by contrasting these two simulations to isolate the physics that is unresolved in the lower resolution simulation, we can identify a key component to driving these sorts of outflows.
        Most importantly, the outflows can't be explained by differences in the macroscopic properties of the galaxy like the total mass, gas(/baryon) fraction, star formation rate, stellar mass, because they are numerically converged between the high and low resolution simulations. 
        Instead, by honing in on the key differing attributes of these simulations we can, hopefully, identify the driving attributes of galaxies which are observed to launch similar outflows in the real universe .

\begin{table}
    \centering
    \begin{tabular}{c|c|c|c}
    Particle group & $N_\mathrm{particles}$ & 3D velocities & $N_\mathrm{fields}$\\ \hline \hline
    \multicolumn{4}{l}{FIRE - m12b\_res7100 - snapshot 600}\\\hline 
        Gas & 8,369,779 (7${\times}$) & \checkmark & 4 \\
        Stars & 1,686,946 (10${\times}$) & \checkmark & 3 \\
        HRDM & 1,490,740 (50${\times}$) & \checkmark & 2 \\
        LRDM & 135,019 (50${\times}$) & \checkmark & 2 \\
        \hline \hline
    \multicolumn{4}{l}{FIRE - m12b\_res57000 - snapshot 600}\\\hline 
        Gas & 6,225,729 & \checkmark & 4 \\
        Stars & 3,264,723 & \checkmark & 3 \\
        HRDM & 932,304 (10${\times}$) & \checkmark & 2 \\
        LRDM & 304,226 (10${\times}$) & \checkmark & 2 \\
        \hline \hline
        \multicolumn{4}{l}{SDSS DR17}\\\hline
        Disks & 311,212 & $\times$ & 6\\
        Ellipticals & 252,715 & $\times$ & 6\\
        Uncertain & 106,795 & $\times$ & 6\\
        Total & 670,722 & $\times$ & 6\\
        \hline \hline
        \multicolumn{4}{l}{Gaia DR3}\\\hline
        no RV & 1,467,744,818 & $\times$ & 2\\
        RV sample & 33,812,18 & \checkmark & 3\\
    \end{tabular}
    \caption{\label{t:FIRE_dataset} the different particle groups, their sizes, whether they have 3D velocities, and the number of scalar fields for each example dataset.}
\end{table}
        
    \subsection{SDSS DR17} \label{s:sdss_example}
        For this example, we use DR17 of the SDSS galaxy survey which can be obtained from the online repository \citep{Meert2017}.
        The Jupyter notebook we use to download the data and generate this instance of Firefly is available at \href{https://github.com/agurvich/SDSS_test}{github.com/agurvich/SDSS\_test}.
        Each of the 670722 galaxies in the sample has RA and Dec sky coordinates along with an estimated redshift. 
        We use \code{astropy} \citep{AstropyCollaboration2013,AstropyCollaboration2018} to convert these into x,y, and z coordinates in 3D space. 
        The \cite{Meert2017} dataset also includes probabilities for basic morphological classifications from the citizen science project Galaxy Zoo for each galaxy, which we use to partition the dataset into three categories \citep{Lintott2008,Lintott2011}.
        We choose to categorize galaxies with $P(\mathrm{disk})\geq 0.5$ and $P(\mathrm{elliptical})<0.5$  as ``likely disks,'' galaxies with $P(\mathrm{elliptical})\geq 0.5$ and $P(\mathrm{disk})<0.5$ as ``likely ellipticals,'' and galaxies with both $P(\mathrm{disk})<0.5$ and $P(\mathrm{elliptical})<0.5$  as ``uncertain.''
        The number of galaxies that fall into each category is listed in Table~\ref{t:FIRE_dataset}.

        In addition to the morphological probabilities, \cite{Meert2017} also include distance modulus, gri apparent magnitudes, extinctions, and k-corrections which we use to calculate absolute magnitudes.
        From these absolute magnitudes we calculate the corresponding luminosities (in units of the solar luminosity) in the same band in the standard way, i.e.: 
        \begin{equation}
            L_b / L_{\odot,b} = 10^{-0.4(M_b - M_\odot)}
        \end{equation}

        We then remap those luminosities to false color RGB by first matching bands to a color channel corresponding to their ordering in wavelength; i.e. (R:i, G:r, B:g).
        For a given color channel/band pair, the final remapping is defined as: 
        
        \begin{equation}
            \mathrm{color~value} = \frac{\log_{10}\left(L_{\mathrm{band}}\right) -\log_{10}\left( L_\mathrm{1\%}\right)}{\log_{10}\left(L_\mathrm{99\%}\right) -\log_{10}\left( L_\mathrm{1\%}\right)}
        \end{equation}
        
        Where $L_\mathrm{1\%}$ and $L_\mathrm{99\%}$ are the 1st and 99th percentile luminosities for the combined list of all three bands.
        We then create a fourth particle group that contains every galaxy and assign them the RGB color corresponding to their gri magnitudes.
        We initially disable this particle group using the PDPP so as not to overlap with the partitioned particle groups but include it so that we can toggle between.
        
        The resulting visualization (shown in Figure \ref{f:full_screengrab}c) reveals a number of interesting features in the SDSS data.
        
        First, we that the nearest and furthest galaxies are all classified as uncertain.
        Enabling the gri magnitude colored particle group (and disabling the morphologically classified particle groups) shows that the nearest galaxies have very low absolute magnitudes (the particles appear nearly black) suggesting that the reason they are classified as uncertain is because they are dimmer (smaller/lower mass) galaxies which are only detected as a result of their (relative) proximity and thus are more likely to be irregular.
        On the other hand, the furthest galaxies are also the most difficult to detect and so are less likely to have obvious features for the citizen scientists to classify.

\begin{figure}
    \centering
    \includegraphics[width=\linewidth]{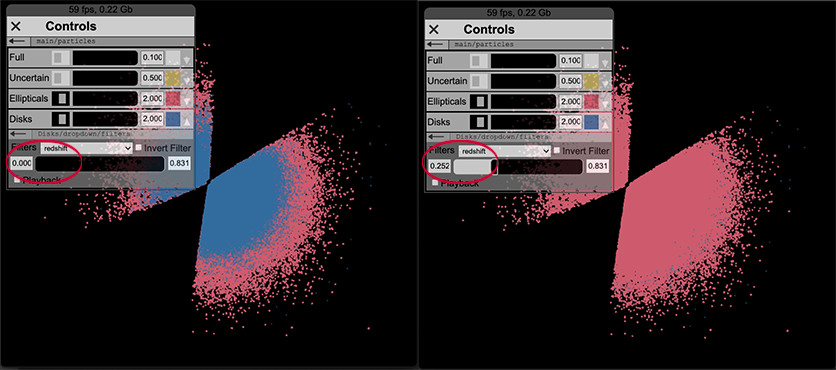}
    \caption{\label{f:sdss_figure} demonstrating the usage of the redshift filter limits to find the maximum extent of disk galaxies in the SDSS dataset, $z\approx 0.25$.}
\end{figure}
        If we switch the view back to the morphological classifications, change the blending mode from additive to normal, and enable the depth buffer checkbox, the pixel color value we see represents the morphology of the galaxy closest to the camera's current perspective along each line of sight.
        This is a valuable viewing mode to differentiate the colors of overlapping particles and the normal+depth blending mode is enabled automatically when the colormap is enabled.
        
        By comparing the frequency of the blue points to red points as a function of distance from the center we can see that with increasing redshift disks become much rarer in the dataset compared to ellipticals.
        We can estimate the maximum redshift the bulk of disk galaxies are contained within by interactively moving the redshift filter minimum handle to show that at a redshift of $z\gtrsim 0.2$, ellipticals dominate the galaxy population (see FIgure \ref{f:sdss_figure}).
        There are likely two effects at play that lead to this relative dirth of disks for redshift $z\gtrsim 0.2$ in the dataset: 1) the physical effect that disks are actually rarer as one looks further back in time (a phenomenon known as ``disk-settling,'' see e.g. \citealt{Kassin2012}) and 2) the observational effect that higher mass galaxies (which are preferentially elliptical) are brighter and therefore can be seen from further away. 
        The ease with which these effects can be illustrated and communicated (e.g., in a public lecture or museum setting) by interactively exploring the dataset, again, demonstrates the power of interactive visualization.
    
    \subsection{Gaia DR3}\label{s:gaia_example}
     
        For our final example, we use the latest data release from Gaia, DR3 \cite{GaiaCollaboration2016,Babusiaux2022}.
        The dataset contains RA, Dec, parallax, proper motions in RA and Dec, radial velocity, BP-RP color, and g-band magnitude. 
        We again use \code{astropy} to convert the RA, Dec and parallax to 3D positions for all stars.  Where radial velocity data is available we also include proper motions and radial velocities to calculate the 6 dimensional cartesian position and velocity phase space. 
        Following this, we use the g-band magnitude to scale the particle radii. 
        Lastly, we use the measured difference in blue band and red band magnitudes (\code{bp-rp}) to colormap the stars.
        
        The full DR3 dataset has 1,467,744,818 sources, of which 33,812,183 have radial velocity measurements which are required for the full 6d phase space position (x,y,z + vx,vy,vz).
        The sheer volume of data for this dataset is more than most web browsers will allow a single tab to use, even if the hardware is available.
        Most browsers will only allow ${\sim}$2-3 GB of data to be stored in memory; the Gaia data is ${\sim}$250 GB in total on disk.
        Thus in order to interactively explore large datasets like Gaia, we implement and employ a progressive rendering scheme which only loads the data from disk which is currently in the camera's view.
        We do this by reorganizing the data such that spatially collocated particles are grouped on disk in files which are indexed by octants of an octree.
        
        Our algorithm for building the octree in Python, as well as walking it and loading the necessary data in Javascript, is described in detail in Appendix \ref{s:octree}.
        In short, we start building the octree by defining a bounding box which contains the 99\% of the particles closest to the center of mass and iteratively sort those particles within the box into octants (and then sub-octants, sub-sub-octants, etc.) with a refinement criterion  based on a (configurable) maximum allowable number of particles.
        To accommodate extremely large datasets (like this example) our implementation does not require the entire dataset ever be loaded into memory at a time and allows multiple worker threads to contribute to building the octree in parallel.  
        Using a single node with 52 cores on the Quest HPC at Northwestern, we built this octree in ${\sim}$ 45 minutes. 
        To determine which data to load in Javascript, we loop through the nodes of the octree and determine: 1) whether any of the vertices are onscreen and 2) the distance from the node center to the camera.
        If the node is onscreen and the distance to the camera is such that the node covers multiple pixels, then the data indexed by the node is queued to be loaded from disk.
        
        Once the data is indexed into an octree and loaded into Firefly (see Figure~\ref{f:full_screengrab}d), we can immediately see interesting features.
        Firstly, there is a dirth of stars along a dark streak which, evidently, traces out the galactic midplane where dust attenuation is strong enough to prevent the detection of any stars along the line of sight. 
        Second, the stars which are along the edges of the galactic midplane have colors that are redder than those stars that are further away owing to the reddening effect of interstellar dust.
        Lastly, by enabling the ``animate velocities'' feature (i.e., extrapolating the positions of particles along the direction of their velocity vectors) we can see that there are (hyper-velocity) stars which move much faster than the galactic average.
        It is also possible to look for co-moving groups of stars to visually identify astrometric binaries and even star clusters and associations.  
        The sheer volume of data makes this last activity challenging through visual inspection alone, but could be facilitated by first identifying candidate groups (e.g.,  by using a friends-of-friends or clump finding algorithm in pre-processing) and adding these as an additional particle group of a different color. 
        
        The Jupyter notebook we use to generate this instance of Firefly is available at \href{https://github.com/ageller/Firefly/blob/main/src/firefly/ntbks/GaiaDR3.ipynb}{the github repo}  and the resulting visualization is available to explore interactively at \href{https://www.alexbgurvi.ch/Firefly}{alexbgurvi.ch/Firefly}.

\section{Performance testing} \label{s:performance_tests}

    Because Firefly runs locally in the browser, its performance is determined by the client's hardware.  
    A Firefly visualization can run on essentially any device that has a WebGL-enabled browser, though performance can vary significantly.
    For instance, a Firefly visualization can be created and viewed locally on a personal laptop.  
    Alternatively, the visualization could be hosted on a powerful HPC, e.g., with many GB of available RAM and a powerful graphics processor.
    As described in \S\ref{s:general_use_cases} above, there are two modes to access the visual content from the HPC server: 1) port forwarding (in which case performance is limited by the client hardware; i.e. the user's laptop) and 2) the \code{/stream} entry point (in which case performance is limited by the HPC hardware and the user's internet connection speed).
    Nonetheless, the most common mode of using Firefly will no doubt rely on the computing resources available on a standard consumer grade laptop.
    
    The general criteria and limitations that govern the performance with which a given dataset is loaded and visualized \textit{interactively} are as follows: 
    
    \begin{itemize}
        \item space on disk that the data occupies (limited by available hard drive space)
        \item amount of RAM required to store the dataset in memory (limited by the amount of RAM installed and possibly further limited by the amount of RAM allowed within the browser)
        \item time to load the data from disk into RAM (limited by the CPU clock speed)
        \item time to render an individual frame (limited by the GPU)
    \end{itemize}
    
    In this section, we provide benchmarks run using a Macbook Air (M1, 2020) running macOS Monterey Version 12.3.1 with 8 GB of RAM purchased in March of 2022. 
    The guidance here is intended to convey confidence that a modern personal computer can run Firefly interactively with a large dataset; one does not need a purpose built workstation. 
    However, each dataset is unique, and the only way to know if a dataset will run well in Firefly is to make an attempt.

\begin{figure}
    \centering
    \includegraphics[width=\linewidth]{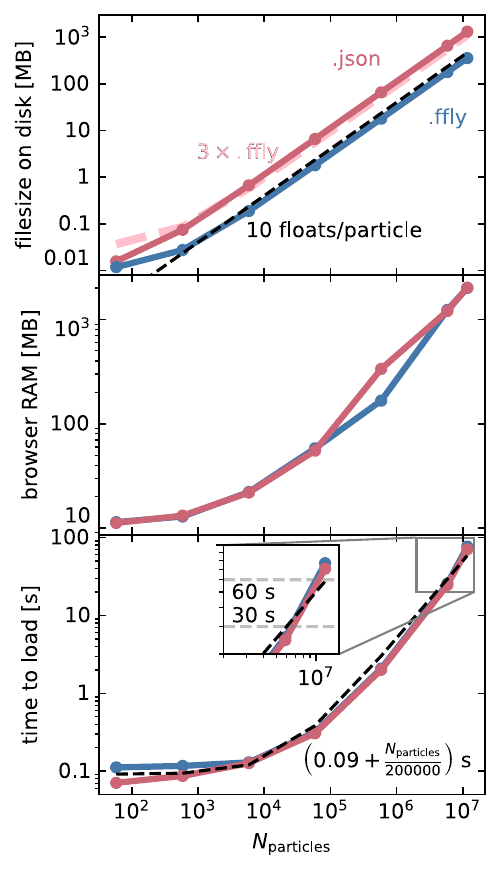}
    \caption{\label{f:perf_startup} Performance benchmarks as a function of the number of particles in a test FIRE dataset (\code{m12b\_res7100}).
    Different numbers of particles are achieved using different levels of decimation.
    Dashed lines are models (defined by the black annotations in each panel) whose parameters are manually fit according to the \code{.ffly} dataset in each panel.
    \textit{Top:} the size of the \json/ and \code{.ffly} files when the dataset is (re)written to disk. 
    \textit{Middle:} the amount of RAM being used by the browser after opening the \json/ and \code{.ffly} files. 
    The size is expected to be larger than on disk as Firefly requires additional color/size/etc. arrays in addition the raw particle data.
    \textit{Bottom:} the time it takes for Firefly to open the \json/ and \code{.ffly} files from disk and initialize the scene.
    In general, we see little difference between the \json/ and \code{.ffly} output types except when it comes to the size on disk, where \json/ is inflated by a factor ${\approx}3\times$, as predicted.
    Finally, a rough rule of thumb for the amount of time it should take to open a dataset is roughly 1 second for every 200000 particles.
    }
\end{figure}
\begin{figure*}
    \centering
    \includegraphics[width=\linewidth]{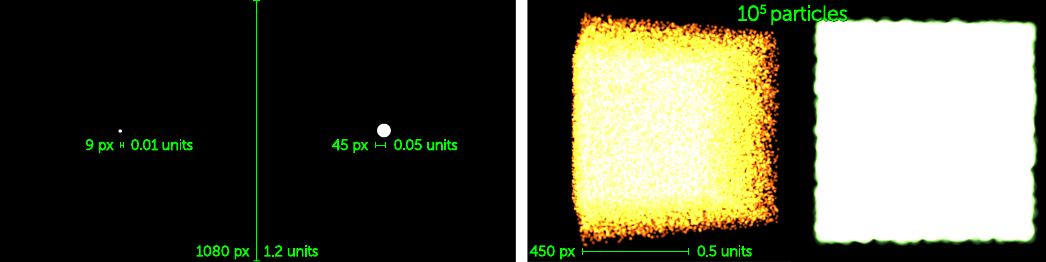}
    \includegraphics[width=\linewidth]{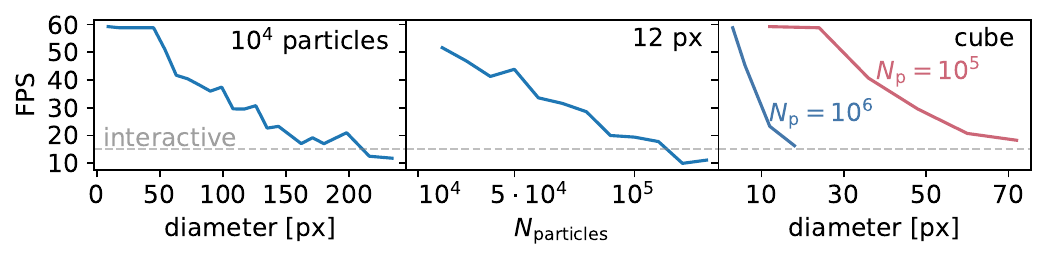}
    \caption{\label{f:perf_fps} performance benchmarks for the framerate of a Firefly visualization as a function of number of overlapping particles.
    In general, the bottleneck for performance during the visualization is how many particles in a pixel are overlapping (and for how many pixels this number is ${\gg}1$).
    \textit{Top:} schematic diagrams of the two tests we perform: 1) placing many particles at the origin (left) and 2) distributing many particles in a cube of fixed size.
    In both cases, we vary both the number of particles and their diameter.
    Scale bars illustrate the mapping between dataspace (in ``units'') and screen space (in pixels) from a fixed camera distance 2 units for consumer grade monitor with a 1920x1080 resolution.
    In all tests, we move the camera along a predefined path orbiting the origin with radius of 2 units.
    \textit{Bottom:} quantitative results of varying the number and diameter of particles for both the overlapping particle (left 2 panels) and cube (right-most panel) setup.
    In both of these extreme cases of particle density, we find that interactive framerates are achievable.}
\end{figure*}

\begin{figure}
    \centering
    \includegraphics[width=\linewidth]{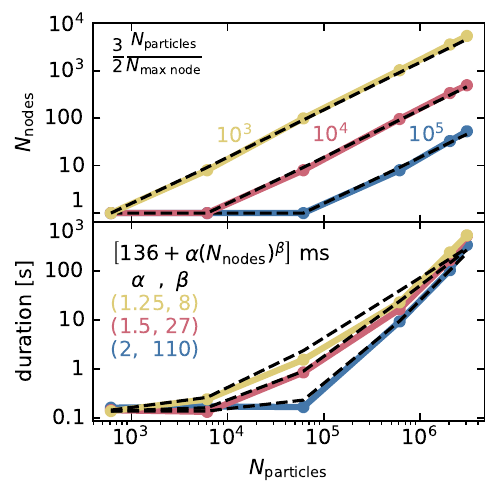}
    \caption{\label{f:perf_octree} Performance benchmarks for pre-formatting sample FIRE data as an octree using the Firefly PDPP. 
    \textit{Top:} the number of nodes in the constructed octree as a function of the number of particles in the dataset and the maximum number of particles in a node before it is refined into child nodes (yellow $10^3$, red $10^4$, and blue $10^5$).
    \textit{Bottom:} the time elapsed constructing the octree. 
    The dashed lines in both panels show fitting functions (defined by the black annotations in each panel) to the data for each $N_\mathrm{max~node}$.
    Generally, octrees with fewer but larger nodes take less total time to generate but appear more ``chunky'' than those with more numerous and smaller nodes.
    }
\end{figure}   

    \subsection{Memory limitations and startup }
        Our first set of benchmarks aim to measure the amount of memory (both on disk and in RAM) required to load in datasets of different sizes, as well as the time required to load the data from disk into RAM.
        
        The size of the dataset on disk can be calculated exactly. 
        In this case we use a dataset from the FIRE simulations and apply different decimations to achieve different numbers of particles.
        In this dataset, we include 3 coordinates ($x,y,z$), 3 velocities ($v_x,v_y,v_z$), and 4 scalar fields (the density, temperature, galactocentric radius, and magnitude of the velocity) for each particle (10 single precision floats per particle). 
        The top panel of Figure~\ref{f:perf_startup} shows that the binary \code{.ffly} format scales exactly as predicted except at the lowest end (where the data volume is comparable to the size of the \code{.ffly} file headers).
        
        The \json/ format is inflated by a factor of ${\sim}3$ by comparison. 
        We can understand this by estimating the \textit{number of characters} that would be required to represent a single precision floating point number since \json/ is a text-based format. 
        The number of digits in the typical string representation of a single precision floating point number is ${\sim}10$.
        We must also count the decimal point and about half of the numbers will have a minus sign, so we add $1.5$ characters to our running count. 
        Additionally, the arrays are comma delimited bringing our final count to ${\sim}12.5$ characters to represent a floating point number in a \json/ format. 
        A single precision floating point number occupies 4 bytes whereas a character requires only 1 byte, thus, we expect \json/ files to be inflated by a factor of ${\sim}12.5/4\approx 3$, as we see empirically.
    
        Note that writing to disk is only necessary if hosting on the internet or if using an octree.
        Otherwise, writing to disk can be totally avoided if the data is sent via Flask instead which does not require writing intermediate files to disk at all.
        
        The space that each particle occupies in RAM once the application is started is less obvious. 
        In addition to the raw data as it exists on disk, there are additional arrays that must be created like: RGBA color arrays (4 additional floats) and particle sizes (1 additional float). 
        The coordinate and velocity arrays must also be copied into special purpose buffers (6 additional floats), as well as the scalar field currently being colormapped (1 float; even if colormapping is disabled this buffer must be filled with 0s).
        As well, copies of subsets of the data are made as filters are applied, colormaps are computed, etc. making it difficult to exactly predict the memory footprint as a function of the number of particles and scalar fields.
        
        Finally, and perhaps only relevant for small datasets, the various Javascript libraries we employ have to be loaded as well (representing a constant offset).
        The middle panel of Figure~\ref{f:perf_startup} measures the memory usage as reported by Chrome's internal memory usage API (\code{window.performance.memory.totalJSHeapSize}).
        In it we see that, as one would expect, there is not a significant difference between whether the data is loaded from \json/ or \code{.ffly} format. 
        
        Lastly, the time required to load the data from disk into RAM will vary substantially between different datasets and computers. 
        In particular, the speed and condition of the RAM, the speed of the CPU, and the speed of the hard drive will all play a role. 
        Any other processes running on the computer will also effectively slow down Firefly's ability to read in data.
        The bottom panel of Figure~\ref{f:perf_startup} shows the measured timings for this specific combination of dataset and hardware both for \json/ and \code{.ffly} formats. 
        In general, the two are indistinguishable except for the very largest datasets. 
        The best-fit timing to the \code{.ffly} data is, as a rule of thumb, 1 second for every 200000 particles in the dataset.
      
    \subsection{Interactivity in the web application} \label{s:particle_count}
        The number of frames rendered per second (FPS) determines whether the app is experienced ``interactively.''
        For our purposes here, we'll define an interactive FPS as being above 15 FPS. 
        In Firefly, the FPS falls steeply when many particles overlap one-another, and even more steeply when this is true for many pixels on the screen.
        The issue being that the hardware accelerated rendering is performed on each pixel independently (and in parallel as threads are available). 
        Thus, when the work required to render an individual pixel increases, it acts as a bottleneck for the application as a whole (note that this is generally true for all graphics rendering processes).
        As you add more particles and the number of overlapping particles increases, or if the size of each of the particles increases, the FPS will drop, eventually below our interactive threshold. 
        
        To give a rough idea of this limitation we consider two idealized scenarios: 1) maximally overlapping particles all placed at the origin (top-left of Figure~\ref{f:perf_fps}) and 2) a cube of uniformly distributed particles between $\pm0.5$ in each coordinate axis centered on the origin (top-right of Figure~\ref{f:perf_fps}). 
        In both scenarios we place the camera on a fixed path circling the origin at a radius $r=2$.
        To automate these tests we use \href{https://pptr.dev/}{Puppeteer}, a \code{Node.js} module that allows users to interact with a basic Chromium web browser using scripted commands. 
        The Puppeteer script we use for these tests is available at \href{https://github.com/ageller/Firefly/blob/main/test/firefly-puppet.js}{the Firefly Github repository}.
        
        In general, the bottom panels of Figure~\ref{f:perf_fps} demonstrate that interactive framerates are easily achievable for moderately sized particles (${\sim} 12$px from this camera distance) for up to $10^5 - 10^6$ overlapping particles.
        Thus by manually adjusting the particle size multiplier as the camera distance increases/decreases, users \textit{should} be able to maintain interactivity without hitting performance bottlenecks except in the most extreme cases of particle density.

    \subsection{Using an octree}\label{s:performance_tests_octree}
        Pre-formatting a dataset as an octree using the PDPP may allow users to avoid some of the limitations described in the previous sections while still visualizing the entire dataset (albeit not all simultaneously). 
        However, there is the obvious upfront time and storage costs of re-writing the data to disk and of processing the data into an octree format.
        
        Octree files are only supported in binary format, so their size on disk will scale the same as in the top panel of Figure~\ref{f:perf_startup}.
        The number of nodes in the octree will determine the level of granularity (or ``chunking'') in the visualization as raw particle data is loaded. 
        In general, having more nodes will produce a visualization that more closely resembles the raw particle data since the nodes will be smaller and faster to load. 
        The maximum number of particles a node can contain before being further refined determines how many total nodes there will be in the octree. 
        The top panel of Figure~\ref{f:perf_octree} shows how the number of nodes scales with the number of particles for three different values of the maximum number of particles per node.
        In all three cases, the number of nodes scales linearly with the number of particles.
        The storage space on disk is constant regardless of the number of nodes except in the limit of small numbers of particles in which case the metadata of the octree structure can approach the total size of the particle dataset, as in the top panel of Figure~\ref{f:perf_startup}. 
        
        The additional time it takes to build octrees with more nodes at fixed number of particles can be significant.
        The bottom panel of Figure~\ref{f:perf_octree} demonstrates that, as one might expect, the average time to build a node, $\alpha$, is longer for larger nodes.
        Also, the total time to build the octree scales more steeply, $\beta$, with larger nodes.
        Despite the larger $\alpha$ and $\beta$ for larger nodes, the total time is still shorter as $N_\mathrm{max~node}$ is increased.
        Thus, we recommend users err on the side of larger nodes in general and we set the default maximum number of particles per node in the PDPP to $10^5$.  Users who wish to visualize their own data using Firefly's octree mode can (and should) experiment with the octree parameters within the PDPP to achieve their desired visual effect within Firefly, using the results of these performance tests as a guide.

\section{Conclusion} \label{s:conclusion}
    
    In this paper we introduce and describe the inner workings of Firefly, a browser-based 3D interactive visualization software for particle data.
    Firefly is lightweight, portable, and performant.
    Viewing a Firefly visualization requires no installation; users need only to visit a URL to immediately explore a pre-configured dataset.
    Because Firefly is browser based, it is automatically compatible with any device or operating system whose internet browser supports WebGL (which is typically true of most popular browsers, e.g., Google Chrome, Mozilla Firefox, Safari, etc.). 
    Firefly is capable of displaying ${\gtrsim}$ 10 million points simultaneously, limited by the amount of RAM available to the browser (which is 2-3 GB/tab in most web browsers).
    This limitation can be circumvented by pre-processing data into an octree which allows Firefly to load on demand only that data which is currently in the camera view. 
    This allows Firefly to visualize datasets with ${\gtrsim}10^9$ particles, like Gaia DR3 (albeit not in its entirety simultaneously). 
    Through the user interface, users can interactively apply filters and colormaps to create new perspectives that can be exported as a screenshot or as a settings file to be imported into another Firefly session (or more computationally expensive rendering tool).
    
    Firefly is bundled with a custom Python data pre-processor (PDPP).
    This PDPP can format user data, customize the UI, host local Firefly servers, and help users create standalone versions of Firefly they can host on the internet. 
    These features allow users to conveniently create, share, and explore streamlined and intuitive Firefly visualizations for different audiences without having to change any of the source code. 
    
    When used locally, Firefly has additional features that are available when it is served by a \flask/-enabled server, which provides a Python-Javascript interface using web-sockets.
    These web-sockets allow users to split the UI and viewer into different windows, explore their scene in VR, stream the rendered image to a client, and pass data to an active Firefly instance without having to write it to disk.
    The last feature is especially useful as it enables users to interactively load and visualize their data seamlessly in a Jupyter notebook. 
    
    The source code to run Firefly locally or to generate new instances of Firefly to host on the web is easily obtained using the \code{pip install firefly} command from a terminal.
    It can also be downloaded and built from its Github repository (\href{https://github.com/ageller/Firefly}{github.com/ageller/Firefly}).
    Detailed Instructions for using Firefly and its PDPP can be found online at \website/ along with a gallery of example Firefly instances and a link to the source code repository.
    
    We also present three examples of applying Firefly to astronomical datasets: 1) the FIRE cosmological simulations, 2) the SDSS galaxy catalog, and 3) the Gaia DR3 dataset. 
    The Python code used to produce these examples are packaged alongside the source code and and particle data in their respective Github repositories.
    Though we have presented Firefly in the context of these specific examples, we emphasize that Firefly can be used to visualize \textit{any} three dimensional dataset, whether those dimensions are spatial or not. 
    Higher dimensional datasets can be visualized using colormaps and filters.
    
    Development of Firefly is ongoing and this paper is linked to the release of the 3.1 version of the code. 
    New features, like greater support for datasets that vary with time, and volume rendering particle data with spatial extent. 
    For an up-to-date list of features and planned changes please consult the project's homepage, \website/.

\section*{Acknowledgments}
This software was based on a prototype by Alessandro Fabretti comissioned by Claude-Andr\'e Faucher-Gigu\`ere and student contributors include Larry Luolei, Nora Linzer, and Mahlet Shiferaw.
The authors would like to thank: Claude-Andr\'e Faucher-Gigu\`ere for his support and guidance; Zachary Hafen for his usage of Firefly which inspired many of the features presented in this paper; them both for their extremely useful comments which improved the quality of this manuscript; Matthew Turk and Cameron Hummels for their guidance and support while developing Firefly; Mariangela Bernardi and Helena Dominguez Sanchez for their helpful advice and guidance in preparing the SDSS test dataset.
ABG was supported by an NSF-GRFP under grant DGE-1842165 and was additionally supported by NSF grants DGE-0948017 and DGE-145000. 
This research was supported in part through the computational resources and staff contributions provided for the Quest high performance computing facility at Northwestern University which is jointly supported by the Office of the Provost, the Office for Research, and Northwestern University Information Technology.
Additional support for student contributors and computational resources came from NSF AST-1715216, AST-2108230, and CAREER award AST-1652522 and a Cottrell Scholar Award to Claude-Andr\'e Faucher-Gigu\`ere.

\bibliographystyle{aasjournal}
\bibliography{bibliography.bib}

\restartappendixnumbering
\appendix
        
\section{How to use a Firefly visualization} \label{s:webapp_features}

    In this section we describe how to navigate around a Firefly visualization within a web browser.  
    In general, to create a new visualization a user first will need to process their data with our Python utilities, which we will describe in the subsequent section.  
    Here we assume that the user already has access to a live Firefly visualization (e.g., one of the examples that we provide on \website/).  
    
    The visual side of Firefly is a primarily Javascript application that uses the \threejs/ library to create a WebGL rendering canvas.
    Along with the WebGL rendering canvas, the \threejs/ library also provides useful primitive objects for representing and manipulating the scene.
    Additionally, we use \kaitaiio/ to build custom Javascript binary data loaders and the \dthreejs/ library to both: 1) load \json/ data/settings files and 2) procedurally create elements of the user interface (UI).
     
    Figure \ref{f:full_screengrab} shows example Firefly scenes.
    The bulk of the screen space is devoted to the rendered scene and a collapsible UI appears on the left side of the screen by default (though it can be clicked and dragged to any position on the screen if desired). 
    Clicking and dragging outside of the UI will move the camera; more information on this is provided in the following subsection.

    \subsection{Manipulating the 3D scene} \label{s:camera_controls}
    
        Users navigate the camera through the visualization using mouse and keyboard inputs.
        The mapping between mouse movements and key presses is split between two control modes: ``trackball controls'' and ``fly controls.''  
        Pressing the space bar, or checking the lock camera check box in the UI, toggles between these two control modes.
        
        Trackball controls, which use only the mouse, are enabled by default. 
        In this camera mode, the view focuses on and orbits a ``camera center'' when users click the left mouse button and drag the mouse. 
        The user can also zoom in and out by scrolling up or down and  pan the camera center by holding the right mouse button and dragging. 
        After moving the camera in trackball controls mode, it will briefly continue moving to provide a sense of inertia.
        
        In fly controls, the primary input device is instead the keyboard, and the camera is moved freely about the scene (without orbiting about a fixed center as with trackball controls).
        Users move forward or backward using the ``W'' and ``S'' keys, left or right with the ``A'' and ``D'' keys, and vertically up or down using the ``R'' and ``F'' keys. 
        Clicking and dragging the mouse up or down (left or right) in this mode will pitch (yaw) the camera.
        Holding the ``shift'' key in combination with any of the above will reduce the speed for fine-adjustment.
        Users can also press the ``+'' or ``-'' keys to increase or decrease the default fly speed (respectively).

        In addition to interactively exploring a dataset using the mouse and keyboard, Firefly has a UI which starts collapsed in the top left corner by default. 
        Clicking the three bars (``x'' that takes its place) will expand (re-collapse) the interface.
        The user interface is organized hierarchically in order to keep the interface compact.
        At the top level, the user interface is split into two categories: 1) \textit{general} controls which affect the application as a whole (\S\ref{s:general_controls}) and 2) any number of \textit{particle group} controls which affect only individual particle groups (\S\ref{s:particle_controls}).
        Users can navigate through the hierarchy by clicking on labeled buttons to move downwards, and clicking on the back arrow to move upwards.
        Users interact with the visualization through buttons, checkboxes, drop-down menus, text entry boxes, one-sided sliders, two-sided sliders, and color pickers.
        Since the UI is procedurally generated using \dthreejs/ to add HTML divs to the DOM, \textit{each section, and their individual elements,} can be disabled independently by adding a unique ``path'' identifier to the \code{GUIExcludeList} (see Appendix~\ref{s:python_pdpp} below).

\begin{figure}
    \centering
    \includegraphics[width=\linewidth]{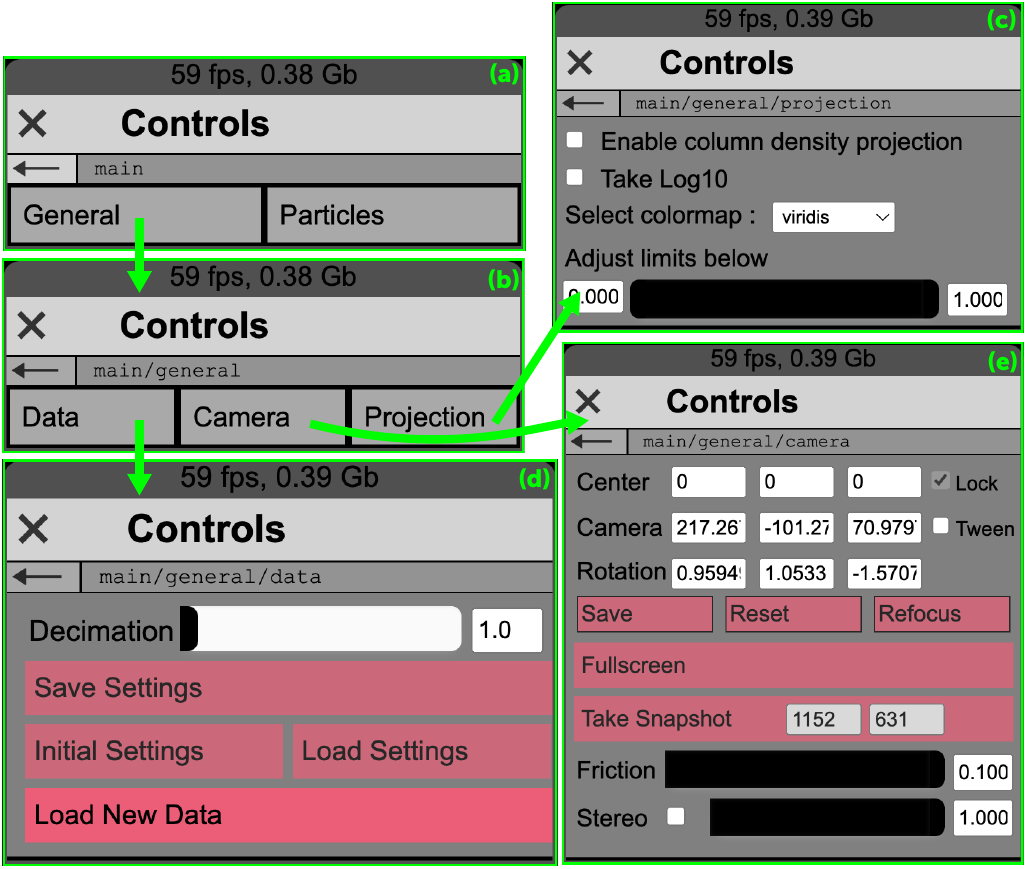}
    \caption{\label{f:app_controls} UI elements that affect the application as a whole. 
    Panes are organized hierarchically and are accessed by clicking labeled buttons (and the backwards arrow in the top left corner) to traverse the tree structure.
    Lime-green arrows connect buttons to the panes they open.}
\end{figure} 

    \subsection{User interface: general controls} \label{s:general_controls}

        The \textit{general} controls, shown in Figure~\ref{f:app_controls}, control global aspects of the visualization and are split into three categories: 1) \textit{data}, which contains controls for adjusting or replacing the current dataset; 2) \textit{camera}, which contains information about the current camera location and controls for modifying the camera view; and 3) \textit{projection}, which allows users to toggle the column density projection mode and adjust the colormapping. 
        
        \subsubsection{General Controls: Data}
            Clicking the ``Data'' button in the \code{main/general} UI pane will reveal the \code{main/general/data} pane (see \appUIref{d}).
            In this pane, there are buttons, sliders, and text entry boxes that allow the user to control various aspects of, or change entirely, the currently viewed data set. 
            The first UI element is a slider and text box combination which applies a global decimation factor to reduce the number of particles on screen in all particle groups (for performance or personal preference). 
            
            The ``Load New Data'' button allows Firefly to directly open new pre-formatted Firefly datasets or to directly convert \texttt{.csv} and \texttt{.hdf5} files.
            This button only appears if the user is running Firefly through a \flask/-enabled server, described in Appendix~\ref{s:flask_backend}.
            In the latter scenario, the files are parsed through the Python utilities described in Appendix~\ref{s:python_pdpp} and sent directly to the running Firefly instance without having to write intermediate Firefly files to disk.  
            For \texttt{.csv} files, Firefly assumes that individual files contain data for a single particle group. 
            Each file must at least contain columns with names of ``x'', ``y'', and ``z.'' 
            For a given \texttt{.hdf5} file, Firefly can accommodate different particle types within a single file if they are split into different HDF5 groups.
            Each particle type must contain a ``Coordinates'' key (pointing to the $x,y,z$ spatial locations of each particle) or ``x'', ``y'', and ``z'' keys.
            We describe how to load one's own data in more detail in Appendix~\ref{s:python_pdpp} below and in \thedocumentation/.
            
            Lastly, the values of the settings which define the current visualization view can be collectively exported or imported through the ``Save'' and ``Load Settings'' buttons. 
            This can be particularly useful if a user has defined a specific view on their data that they want to be able to reproduce later and/or share with a colleague.
            The settings can also be reset to their initial values from the application startup using the ``Initial Settings'' button.

        \subsubsection{General Controls: Camera}
            Clicking the ``Camera'' button in the \code{main/general} UI pane will reveal the \code{main/general/camera} pane (see \appUIref{e}).
            In this UI pane, the camera's position and orientation are printed in read only text boxes.
            This may be useful if, for example, a user desires to copy the camera settings into another application that will perform more computationally intensive rendering for publication.
            There are also two checkboxes, one labeled ``Lock'' which toggles the camera control mode between fly and trackball (see Appendix~\ref{s:camera_controls} above) controls (just as the space bar does), and a second, labeled ``Tween,'' which will move the camera along a pre-defined path, according to keyframe locations and durations defined in the tween file (\code{tweenParams.json} by default).
            This check box only appears if there is an existing \code{TweenParams.json} file linked to the dataset (see Appendix~\ref{s:python_pdpp} below).
            
            Next, there are three buttons for saving the camera position, resetting the camera to the most recently saved camera position, and refocusing the camera.
            The first two buttons allow users to store a saved camera position and then reset the view to that saved camera position at any time. If the reset camera position button is pressed before the save camera position button then the camera is reset to the original position from initialization.
            The refocus button reorients the camera to face the origin. 
            
            In trackball controls, the ``Friction'' slider allows a user to change how quickly the camera decelerates after movement with smaller values allowing the camera to move for longer (with 0 making the camera continue to move forever). 
            In fly controls, the same slider determines the speed of motion (in addition to the ``+'' and ``-'' keys mentioned above). 
            
            The camera view can also be horizontally duplicated and slightly offset to produce a stereoscopic effect that modern 3D televisions and monitors can overlay to make 3D scenes with polarized glasses.
            The ``Stereo'' checkbox will enable stereo mode, and the adjacent slider controls the strength of the 3D effect by changing the stereo separation.

            The canvas can also be expanded to take up the full screen area using the ``full screen'' button when the application is not already in Full Screen mode.
            Additionally, it is also possible to export static images by clicking the ``Take snapshot'' button. 
            The default resolution of these images depends on the user's computer monitor and the size of the browser window, but the resolution can be specified manually in the text entry boxes within the button to produce large format images for print or publication if so desired. 

        \subsubsection{General Controls: Projection}
            Clicking the ``Projection'' button in the \code{main/general} UI pane will reveal the \code{main/general/projection} pane  (see \appUIref{c}).
            In this pane, there are checkboxes, sliders, and drop-down menus which allow the user to enable and adjust Firefly's column density projection mode. 
            
            When column density mode is enabled particles are first rendered to a texture buffer rather than directly to the canvas in order to sum up the number of particles within a given pixel.
            This number is then used to draw a color from the colormap selected from the drop-down menu in the UI.  
            There is also a checkbox to first take the $\log_{10}$ of the particle count before applying the colormap if so desired. 
            Lastly, there is a two-handled slider which controls the colormap limits.
            
\begin{figure}
    \centering
    \includegraphics[width=\linewidth]{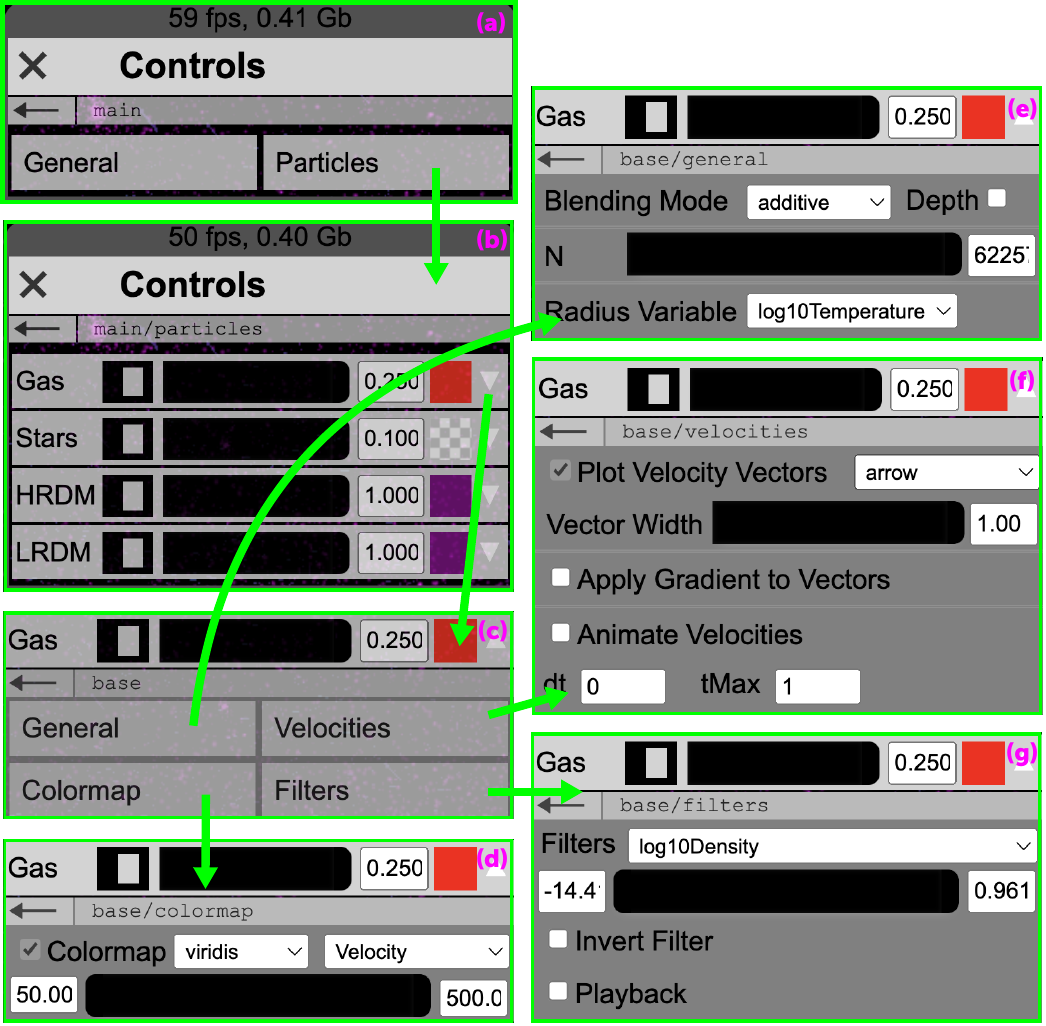}
    \caption{\label{f:particle_controls} UI elements that affect the individual particle groups independently.  The format of this figure is analogous to that of Figure~\ref{f:app_controls}.
    }
\end{figure}

    \subsection{User interface: particle controls} \label{s:particle_controls}
    
        Within a dataset, it is common for there to be different categories or types of data. For example, in cosmological hydrodynamic simulations, there are often at least gas, star, and dark matter particles. 
        In Firefly these different categories are represented as \textit{particle groups}.
        More broadly, particle groups are subsets of data that will share certain characteristics in the visualization like size, shape, and color.
        
        Clicking the ``Particles'' button in the \code{main} pane (\pgUIref{a}) will reveal the \code{main/particles} pane (\pgUIref{b}).
        Figure~\ref{f:particle_controls} illustrates how the various (sub-)panes of the particle UI are organized hierarchically. 
        Each particle group has its own pane in the \code{main/particles} window whose controls affect only that particle group.
        The sub-panes for each particle group are structured identically but only show the relevant fields/limits/etc. corresponding to that particle group. 
        Clicking the downward facing arrow on the right edge of the particle group's row in the \code{main/particles} pane will expand that particle group's \code{base} UI pane and reveal the: general, velocities, filters, and colormap sub-panes (see \pgUIref{c}).
        The downward facing arrow will transform into an upward facing arrow which, when clicked, will re-collapse the \code{base} pane.
        There are three basic controls for every particle group, which are visible regardless of whether the \code{base} pane is expanded: 1) a switch to toggle visibility, 2) a slider+text box to change the particle radius scale factor, and 3) a color picker to change the particle group's color.
        
        \subsubsection{Particle Controls: General}
            Clicking the ``General'' button in the \code{base} particle UI pane will reveal the \code{base/general} pane (see \pgUIref{e}).
            In the first row is a ``Blending Mode'' drop-down menu and ``Depth'' checkbox. 
            The blending mode menu allows users to adjust how the colors of overlapping particles are blended.
            The default mode, additive, adds the RGB values after multiplying by opacity.
            Other modes include ``normal'' (as additive but ignoring opacity), ``subtractive'' (takes the difference of the RGB values).
            The depth checkbox enables the ``depth buffer'', which will render the particles in order from farthest to nearest to the camera; this can be useful when combined with ``normal" blending to show the color of the nearest particle to the camera for each pixel (important when using a colormap).
            Next is the ``N'' slider, which can be used to set the maximum number of particles shown.
            The ``N'' slider is analogous to the decimation slider in the \code{main/general/data} UI pane (\appUIref{d}) but for individual particle groups.
            It can be used to downsample the visualization in order to get a more coarse-grained view without having to create separate input files.
            If the input data for this particle group has scalar fields flagged as being allowed to scale the particle radii (see Appendix~\ref{s:python_pdpp} below), an additional dropdown will be available to change the radius scaling variable for the particle group.
            This is set to ``None'' by default, i.e., all particles in that group are the same size.
            
        \subsubsection{Particle controls: velocities}
            If the dataset includes velocity data then the ``Velocities'' button will be shown in the \code{base} pane. 
            Clicking the ``Velocities'' button will reveal the \code{base/velocities} pane (see \pgUIref{f}).
            Within this pane, there is a checkbox to represent each particle by its velocity vector, rather than a point, a dropdown menu to choose the velocity vector style (from lines, triangles, or arrows), and a slider controlling the width of the vectors.
            The ``Apply Gradient to Vectors'' checkbox toggles whether vectors are colored along their length from the particle group's color (at the tail of the vector) to white (at the head of the vector) to give an impression of directionality.
            Lastly, the velocity ``animate'' controls will linearly extrapolate the particle positions in the direction of their velocity vectors. 
            The ``dt'' and ``tMax'' text boxes control the speed at which the particle positions are extrapolated and the maximum animation duration after which the coordinates are reset to their original position.
            This extrapolation is performed within the graphics shader and therefore adds negligible overhead even for large data sets.
        
        \subsubsection{Particle controls: colormap}
            If the dataset includes scalar field values that have been flagged to colormap, then the ``Colormap'' button will appear in the \code{base} particle UI pane.
            Clicking will reveal the \code{base/colormap} pane (see \pgUIref{d}).
            The ``Colormap'' checkbox toggles the color of each particle in the group between the fixed color for the group and the color corresponding to the particle's value in the selected scalar field array.
            The user can choose which colormap to use and which scalar field to colormap by using adjacent dropdown menus. 
            When colormapping is enabled an additional tab is added to the side of the user interface that contains a colorbar mapping between the scalar field values and the corresponding colors.
            This colorbar is updated automatically if the user changes any of the colormap settings.  
            
            The available colormaps are defined in the \code{colormaps.jpg} file in the \code{firefly/src} directory (mostly drawing colormaps from Python's \matplotlib/ library, \citealt{Hunter2007}).
            This file contains rows of different colormaps whose row index corresponds to their position in a list defined within the \code{colormap\_names.json} file, which is read when Firefly is initialized.
            Replacing these files can allow users to apply custom colormaps if so desired.
            
            The lower and upper limits on the colormap in the two-sided slider are set to the minimum and maximum values for the given field by default, but can also be specified in the settings file.
            Firefly stores the colormap limits for each field separately, so when the user switches between fields the limits will automatically change to those corresponding to the newly selected scalar field. 
            
        \subsubsection{Particle controls: filters}
            If the dataset includes scalar field values that have been flagged to filter by, then the ``Filters'' button will appear in the \code{base} particle UI pane.
            Clicking will reveal the \code{base/filters} pane (see \pgUIref{g}).
            Filters apply additively and can be stacked in order to restrict the visualization to a specific region of phase space.
            Filters and their limits are only applied on a per-particle-group basis even if different particle groups share the same filterable scalar field.
            
            Analogously to the colormap pane, there is a dropdown menu containing the field names that are flagged for filtering and the minimum and maximum limits within the two-sided slider are set to the minimum and maximum field values by default (but can also be specified in the settings file). 
            There are also two checkboxes: one for inverting the filter to exclude particles whose scalar field values fall within the limits and one for enabling filter ``playback'' mode. 
            Playback mode will automatically shift the selected region of the two-sided filter slider along the length of the filter limits. 
            Playback allows users to quickly explore isocontours of specific variables without having to manually move the filter handles themselves.

\section{The Python data pre-processor (PDPP)}\label{s:python_pdpp}
    In addition to the core visualization functions described in Appendix~\ref{s:webapp_features}, Firefly includes a Python interface to facilitate: 1) converting particle data to properly formatted Firefly files, 2) customizing the user interface and startup settings of a Firefly visualization, and 3) hosting local and remote Firefly servers.
    To create a new Firefly visualization or host a local Firefly server, users must first obtain a copy of the source code and install the PDPP.
    The development version of the code is available from its \href{https://github.com/ageller/Firefly}{Github repository} while the latest stable version can be obtained from the Python Package Index (PYPI) and automatically installed using the \code{pip install firefly} command.
    Example Jupyter Notebooks (along with YouTube video tutorials) demonstrating the usage of the PDPP are available in \href{http://www.alexbgurvi.ch/Firefly/docs/build/html/search.html?q=Tutorial+notebook&check_keywords=yes&area=default#}{the documentation.}
    
    \subsection{Formatting particle data with the PDPP}
        Users convert their data into Firefly files in either ASCII \json/ (inefficient but portable) or binary \code{.ffly} (efficient but only interpretable by Firefly) containing the coordinate, velocity, and any additional scalar field data associated with each point. 
        \ffly/ files are read using \kaitaiio/, a flexible framework which allows developers to semantically describe their data using \code{.ksy} configuration files and produce data loading routines for any number of programming languages.
        We provide the \code{.ksy} files we used to produce the Javascript files for loading \ffly/ files alongside the source code of Firefly (though most users should not need to use them unless they intend to modify the \ffly/ file format).
        
        Whether \json/ or \ffly/, each particle group is broken up into chunks across a set of files, rather than being saved as a single large file, to improve performance when loading data into the web application.
        We find that 10,000 entries per file is a good size for chunking but users may find that performance varies depending on the particular system they are running Firefly on and are encouraged to experiment with different chunking (using the \code{n\_particles\_per\_file} attribute of the \docref{data\_reader/reader.html\#the-reader-class}{firefly.Reader} class).
        Each file must be listed in the manifest \code{filenames.json} file, which is populated automatically when data is loaded into a  \docref{data\_reader/reader.html\#the-reader-class}{firefly.Reader} instance, so that Firefly knows to load it in.
        Manifest files must be listed in the startup file, \code{startup.json}, which is automatically generated by the PDPP when new datasets are created.
        
        As in the web application, the PDPP organizes data into separate \textit{particle groups}. 
        We provide the \particlegroup/ class, which holds the data and performs basic data validation.
        Each particle group must have a set of coordinates which represent the  x, y, and z locations of each particle in the web application. 
        They can also optionally contain velocity and RGBA color arrays, along with any number of scalar field arrays to use to filter, colormap, or scale the radius of the corresponding points.
        These optional arrays are expected to share indices with the coordinate data and so must have an entry for every particle within a particle group.
        Once data is ingested into a \particlegroup/ instance, the \code{.writeToDisk} method can produce Firefly files. 
        For more information, examples, and tutorials see \thedocumentation{data\_reader/particle\_group.html}.
        
    \subsection{Customizing the UI and startup settings} \label{s:settings_pdpp}
    
        The user can customize the UI to start the application from a specific camera location, set filter/colormap limits, or even disable any UI element using a \code{settings.json} file.
        This allows users to produce instantly shareable visualizations that start from the exact perspective they desire and to tailor their visualizations to different audiences by restricting the level of interactivity available.
        In addition to describing the state of the web application at startup, settings files can also be imported through the ``Load Settings'' button of the UI. 
        Likewise, settings files can be created by clicking the ``Save Settings'' button in the UI.
        
        However, most users will find that it is easiest to generate settings files by hand using the \docref{data\_reader/reader.html\#the-settings-class}{firefly.Settings} class.
        The \docref{data\_reader/reader.html\#the-settings-class}{firefly.Settings} class is implemented as a restricted dictionary with key validation, raising an error if an invalid setting is accessed and providing a best guess alternative key.
        Settings are organized into two broad categories: those that affect the entire application and those that affect individual particle groups. 
        For settings that affect individual particle groups, we use a nested dictionary structure where the particle group's \code{UIname} indexes the value for that setting (e.g. \code{settings[`sizeMult'][`Gas']=1}).
        Within these two broad categories each is split into a handful of sub-categories: startup, UI, window, and camera for the global settings; startup, UI, filter, colormap, and velocity for the per-particle-group settings.
        All settings can be accessed from the \docref{data\_reader/reader.html\#the-settings-class}{firefly.Settings} instance.
        
        To disable elements of the UI, users can add unique identifying paths, which are case insensitive, to the \code{GUIExcludeList} in their \code{settings.json} file.
        Paths are structured hierarchically, e.g., the path for \code{main/general/data/decimation} would disable the decimation slider in the \code{main/general/data} pane whereas \code{main/general/data} would hide the \code{data} button on the \code{main/general} pane and \code{main/general} would hide the general button on the \code{main} pane. 
        If users would prefer to disable the UI entirely, then the \code{UI} key in the \code{settings.json} should be set to False.
        Paths for per-particle-group panes begin with the name of the particle group, e.g. \code{Gas/onoff} and \code{Gas/dropdown/velocities/velocityCheckBox}.
        
        Users can also pre-define camera paths using a \code{tweenParams.json} file generated by a \docref{data\_reader/reader.html\#the-tweenparams-class}{firefly.tweenParams} instance.
        Camera paths are defined by a sequence of key frame camera coordinates that are linearly interpolated between for a specified duration between each frame.
        Once the last keyframe is reached the path will wrap around and loop through the keyframe list again until tweening is disabled (by unchecking the tween checkbox in the \code{main/general/camera} pane). 
        
        Both \code{settings.json} and \code{tweenParams.json} files can have any name so long as they are correctly referenced in the manifest file (i.e., they are attached to a \docref{data\_reader/reader.html}{firefly.Reader} instance).

    \subsection{Hosting local and remote Firefly servers} \label{s:flask_backend}

        To explore a custom instance of Firefly, a user must host these files (along with the Firefly source code) either on the internet (as is done for the examples in the gallery at \href{http://alexbgurvi.ch/Firefly}{alexbgurvi.ch/Firefly}) or locally using the \code{http} module in the Python 3 standard library (invoked as \code{python -m http.server} from the command line while located inside the Firefly source directory).
        Users can also host the files locally using \flask/, a Python backend that effectively connects the browser to a Python interpreter and enables numerous additional Firefly features, including the ability to split the visualization and user interface into different browser windows, stream the visualization from one computer to another, explore data in virtual reality and to pass data directly to Firefly without saving Firefly files to disk (which can take-up significant hard-drive space). 
        These features are accessed by visiting different urls in a web browser (which we refer to as ``entry points,'' as described in Appendix~\ref{ss:entrypoint}).
    
        The PDPP offers multiple ways to launch \flask/ servers: 1) the \docref{reference/api/server/firefly.server.spawnFireflyServer.html\#firefly.server.spawnFireflyServer}{spawnFireflyServer} function which can be used from in a Python script or Jupyter notebook and 2) the \code{firefly} terminal command which is added to the user's path when Firefly is installed (e.g., via \code{pip}).
        It also includes a convenient interface for POST'ing new data using the \docref{reference/api/classes/firefly.data\_reader.Reader.html\#firefly.data\_reader.Reader.sendDataViaFlask}{Reader.sendDataViaFlask} method.
        
        Lastly, the PDPP makes it convenient and easy to host new instances of firefly on the internet using the \docref{reference/api/classes/firefly.data\_reader.Reader.html\#firefly.data\_reader.Reader.copyFireflySourceToTarget}{Reader.copyFireflySourceToTarget} method.
        By default, \docref{reference/api/classes/firefly.data\_reader.Reader.html\#firefly.data\_reader.Reader.copyFireflySourceToTarget}{Reader.copyFireflySourceToTarget} will copy only the necessary Javascript, data, and settings files to the specified path.
        However, if users create a GitHub OAuth authentication token, save it to their computer as a text file, and pass the path to that text file, a new GitHub repository can be automatically created with GitHub pages enabled. 
        Thus, with one line users can immediately push their data to the internet and share the url with a collaborator without extensive background knowledge of web hosting.
        More details can be found \docref{data_reader/multiple_datasets.html?\#with-separate-firefly-source-directories}{in the documentation}.
    
    \subsection{Embedding Firefly within a Jupyter notebook\label{ss:jupyter}}
        Jupyter notebooks are powerful engines for interactively analyzing and interpreting data.
        In light of their growing popularity, we have also included a convenient interface for launching Firefly servers and embedding Firefly visualizations all within the confines of a Jupyter notebook. 
        To take advantage of this feature users need only:
        
        \begin{enumerate}
            \item launch a Firefly server as a background subprocess using the \docref{reference/api/server/firefly.server.spawnFireflyServer.html\#firefly.server.spawnFireflyServer}{spawnFireflyServer} function
            \item access the local url from within the Jupyter notebook using an IFrame
            \item explore the data from within the IFrame
            \item (optional) interactively change the dataset using the \docref{reference/api/classes/firefly.data\_reader.Reader.html\#firefly.data\_reader.Reader.sendDataViaFlask}{sendDataViaFlask} method
            \item close the server when no longer required using the \docref{reference/api/server/firefly.server.spawnFireflyServer.html\#firefly.server.closeAllFireflyServers}{closeAllFireflyServers}
        \end{enumerate}
        
        More detailed instructions and examples are available in \thedocumentation{server.html\#using-firefly-from-within-a-jupyter-notebook}.       

\section{The Javascript application} \label{s:under_the_hood} 
\begin{figure}
    \centering
    \includegraphics[width=\linewidth]{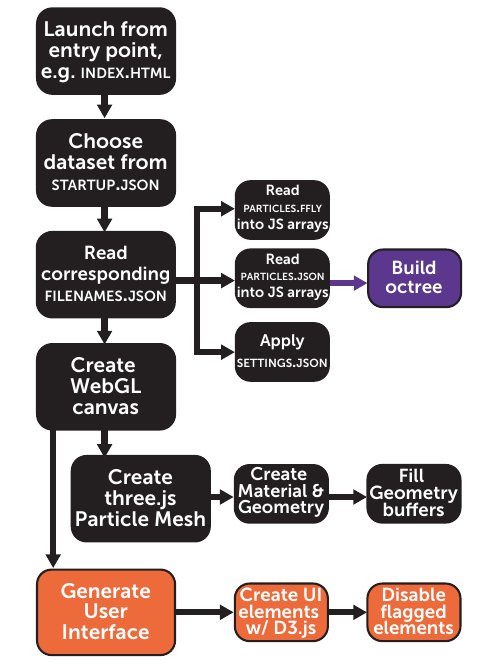}
    \caption{\label{f:startup_diagram} An illustration of the Firefly web application control flow at initialization.
    The steps outlined here are executed only once, in contrast to the components of the interactive render loop which are repeated many times a second (illustrated in Figure~\ref{f:combined_diagram}).
    Major steps are in larger boxes and control flow proceeds from top to bottom.
    Sub-steps are in smaller boxes and proceed horizontally from left to right.
    Colors indicate which aspect of the application handles the step:  viewer in black, UI in orange, and the optional octree build in purple (Appendix~\ref{s:octree}). 
    Major steps are executed sequentially, except for ``Generating the User Interface'' and ``Creating \threejs/ Particle Mesh'', which occur in parallel.
    }
\end{figure}

    Figures \ref{f:startup_diagram} and \ref{f:combined_diagram} illustrate the control flow of the web application from entering a Firefly url to the interactive rendering loop.
    We describe each of these steps in detail below but, in brief, these steps are:
    
    \begin{enumerate}
        \item First, a user visits one of a Firefly server's ``entry points'' (\S\ref{ss:entrypoint}) to initialize the web application in the browser tab's Javascript interpreter.
        \item Then the startup file, \code{startup.json}, is read to determine which dataset to load.   
        If multiple datasets are listed in the startup file the user is presented with a dropdown list to select from. 
        \item Next, the settings and particle data files specified in the manifest file, \code{filenames.json}, are opened using \dthreejs/, and/or \kaitaiio/ where binary particle data in \code{.ffly} format is specified. 
        \item A \threejs/ WebGL renderer is initialized and resized to fit the window. 
        The renderer's keyboard/mouse controls mode and camera properties are set according to the input settings.
        \item \threejs/ material and geometry objects corresponding to each particle group are created and the geometry buffers are filled with coordinate, velocity (if available), and colormap (if enabled) data.
        \item The UI is generated (using \dthreejs/, described in Appendix~\ref{s:webapp_features}) and the settings for filters, particle sizes, etc. from the input settings are applied. 
        \item After the GUI and viewer are initialized, the application enters the visualization render loop which continuously checks the UI and updates the corresponding values in the \threejs/ material objects and geometry buffer.
        \end{enumerate}
        
    \subsection{Start-up and initialization}
        Unless a new dataset is loaded through the user interface, steps 1-6 above (i.e., excluding the render loop) are only performed once when Firefly is initialized.
        These steps configure the application, load in data, and create the objects necessary to visualize the dataset. 
        
        \subsubsection{Startup: visiting an entry point}\label{ss:entrypoint}
             
            An entry point, in this context, is a url that is accessed to initialize the web application. 
            There are four possible entry points for Firefly: 
            \begin{enumerate}
                \item \code{/index} 
                \item \code{/combined} 
                \item \code{/gui} \& \code{/viewer}
                \item \code{/data\_input}
                \item \code{/stream}
                \item \code{/VR}
            \end{enumerate}
            
            If Firefly is being hosted over the internet or using a local \code{http} server, then only the \code{/index.html} entry point will work.
            For instance, on a local server, one would point their browser to \code{localhost:8000/index} to access the first entry point.  
            The other entry points require \flask/.
            If no entry point is specified \code{/index} is implied.
            Once the Firefly source files have been delivered by the server to the client browser, the application is launched into the mode corresponding to the entry point.

            The \code{/combined} entry point (which shares an alias with the \code{/index} and blank \code{/} entry points) shows the default Firefly web application with user interface and viewer combined in one browser tab.  The \code{/gui} and \code{/viewer} entry point pair allows a user to split the user interface and visualization view between two separate windows (or devices on the same network).
            By connecting a primary device like a computer with a high-end graphics card connected to a large screen to the \code{/viewer} entry point, and a second device like an iPad or smartphone to the \code{/gui} entry point, a user can manipulate the visualization remotely in a manner ideal for presentations or public outreach activities.  
            POST'ing Firefly formatted data to the \code{/data\_input} entry point  updates the data currently being viewed without having to write it to disk. 
            The \code{/stream} entry point displays a rasterized version of the visualization scene allowing users to render the Firefly scene on a powerful workstation and view it from a less powerful computer.
            Lastly, the \code{/VR} entry point is experimental and can be used to view Firefly with a VR headset (e.g., Google cardboard) with limitted navigational controls.
            For additional details on how to use \flask/, and which features require \flask/, see \thedocumentation{data\_reader/flask.html\#flask}.
        
        \subsubsection{Startup: selecting a dataset}
            The first step performed by Firefly is to look for input data to visualize.
            The input dataset is specified by the startup file, \code{startup.json} in the \code{Firefly/data} directory. 
            The contents of \code{startup.json} are a dictionary mapping of key-value pairs with numbered indices starting at 0 as keys mapped to the relative path of different dataset directories with respect to the \code{static} directory.
            For browser security purposes, all files loaded by the application \textit{must} be accessible as sub-directories of the \code{static} directory in the source code (e.g, within the \code{static/data} directory). 
            If a user prefers to store their data elsewhere in their computer, one way to achieve this for local servers is to create a symbolic link from the data's location on disk to the \code{static/data} directory (which the PDPP will do for users automatically).
        
            An example startup file might look like: 
        
    \begin{lstlisting}
    {0:`data/snapshot_600',
     1:`data/snapshot_599',
     2:`data/snapshot_598'}
    \end{lstlisting}
            
            If only a single key-value pair exists in the \code{startup.json} file, then Firefly will automatically begin loading the data in that directory.
            If multiple entries are provided (as in the example above), Firefly will prompt the user to choose which data set to load.
            If the \code{startup.json} file does not exist then Firefly will prompt the user to select a directory manually using a file browser. 
            
            Each directory provided in \code{startup.json} must contain a manifest file, \code{filenames.json}, that identifies the filenames of the settings and particle data files within the directory.
            This is created automatically by the PDPP. 
            
        \subsubsection{Startup: loading settings and particle data}
            Once the data directory is identified, Firefly attempts to load the data listed in the \code{filenames.json} manifest file.
            The manifest file includes entries for the settings file (\json/ extension) and every file containing particle data (\json/ and/or \code{.ffly} extensions).
            If Firefly does not find a \code{filenames.json} manifest file and it is not being served from a \flask/-enabled server, it will raise a Javascript alert in the browser indicating it cannot load the data.
            If Firefly is served from a \flask/-enabled server, it will attempt to load every \code{.csv} or \code{.hdf5} file contained in the directory chosen from the \code{startup.json} file using the PDPP.
            If no \code{.csv} or \code{.hdf5} files are found, the same error is raised.
            
\begin{figure*}
    \centering
    \includegraphics[width=\linewidth]{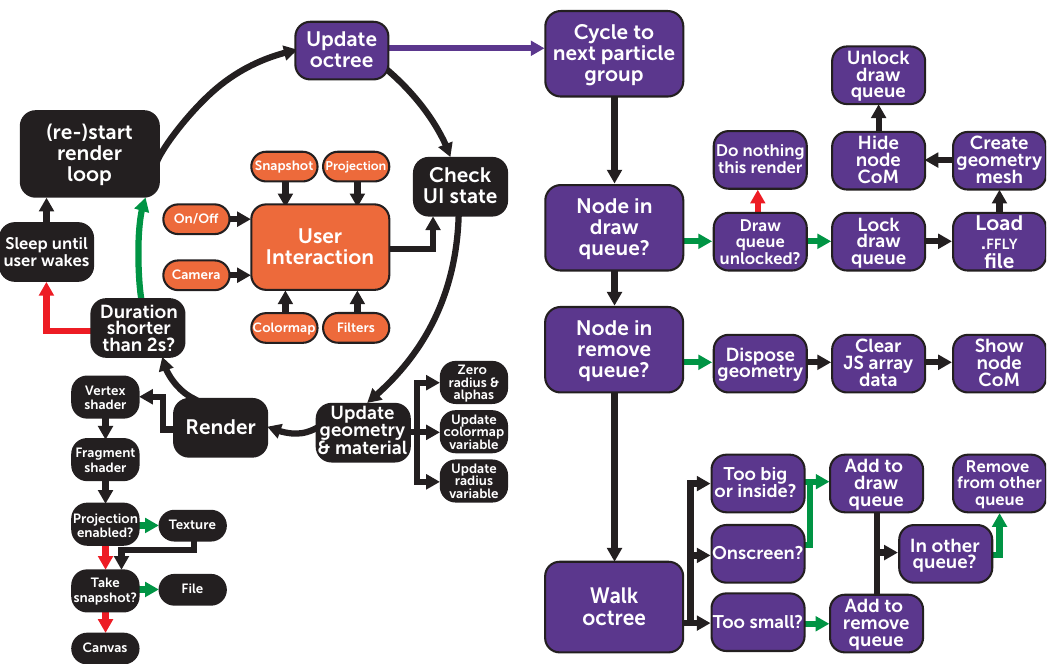}
    \caption{
        \label{f:combined_diagram}
        An illustration of the Firefly render loop control flow.
        The left shows a view of the overarching render loop flow while the right shows a zoomed-in view of the ``Update octree'' step (described in Appendix~\ref{s:octree}).
        Like Figure~\ref{f:startup_diagram}, major steps are in larger boxes and sub-steps are in smaller boxes.
        Colors indicate which aspect of the application handles the step: viewer in black,  UI in orange, and the optional octree steps in purple. 
        Colored arrows indicate conditional branch outcomes where green indicates \code{true} and red indicates \code{false}.
        }
\end{figure*}

            The manifest file also allows users to break up large particle datasets into multiple files.
            There are several advantages to using multiple smaller files rather than using a single large file (which may crash some browsers in their attempt to load it).
            Separating the particles into a series of smaller files allows us to provide a well-defined status bar, representing the fraction of particle files that have been opened, to be presented to the user while the data is loading.
            Lastly, it is common for web-hosting services (like Github pages, a popular and free web-hosting service) to have strict individual file size limits; separating the data into multiple files, each under the size limit, can allow the user to host these data online more easily. 
            
            Once the data is loaded, the settings are imported from the settings \json/ file using \dthreejs/.  These settings are used to set the default values of the visualization, e.g., the filter limits, colormap attributes, particle sizes, which aspects of the user interface to include, etc.

        \subsubsection{Startup: creating the renderer}
            The rendering canvas is a \threejs/ WebGL renderer.
            If stereo mode is requested at startup (within the settings file), this is set now. Otherwise the user can toggle into stereo mode through the user interface. 
            
            The camera is initialized, and attached to either the trackball or fly controls control mode (see Appendix~\ref{s:camera_controls}).
            The render distance is effectively infinite by default but only particles that are large enough to cover multiple pixels are drawn.
        
        \subsubsection{Startup: making material and geometry objects} 
        
            After the UI is generated and the canvas is initialized, Firefly creates a \threejs/ point cloud object for each particle group and copies the data into its memory buffers.
            Each point has, at minimum, a three-dimensional coordinate, size, and opacity.
            If per-particle RGBA colors or a colormappable field are provided then corresponding memory buffers are initialized and filled with the corresponding values.
            If velocity vectors are provided, then each velocity's magnitude is rescaled to the interval [0,1] defined by the minimum and maximum velocities of that particle group.
            The unit vector pointing along the direction of velocity and the new normalization are stored in a flattened $N_\mathrm{part}\times4$ array and passed to a velocity vector geometry buffer.
            Otherwise, dummy buffers are initialized containing all zeroes because the fragment and vertex shaders require them.
        
        \subsubsection{Startup: generating the UI}
            Firefly uses \dthreejs/ to generate the DOM elements which make up the UI dynamically upon loading in the data and settings files.
            Within the settings file, the user can specify which of the controls to exclude from the UI using the \code{GUIExcludeList} setting (see Appendix~\ref{s:settings_pdpp}).
            Additionally, individual particle group panels are automatically customized according to the available data in each particle group (e.g., the velocity vector controls are not shown as an option if velocities were not provided with the data).
            
            Many of the elements of the UI shown in Figures \ref{f:app_controls} and \ref{f:particle_controls} are generated with ``vanilla'' HTML+CSS+JS code, though some use specialized Javascript libraries. 
            Specifically: 1) the one- and two-sided sliders used for adjusting floating point parameters use the \href{https://refreshless.com/nouislider/}{noUiSlider library}, 2) we use the \href{https://refreshless.com/wnumb/}{wNumb library} to format and interpret many of the numbers displayed in the UI (particularly in connection to the sliders), and 3) the color picker uses the \href{https://bgrins.github.io/spectrum/}{spectrum library}, which itself requires \href{https://jquery.com/}{jQuery}.
            
    \subsection{The main render loop} \label{s:render_loop}

        Once all of the data and settings files are loaded, the application enters the render loop.
        The render loop uses the \code{requestAnimationFrame} function to repeatedly call the \code{animate} function. 
        Using \code{requestAnimationFrame} (rather than, say, a \code{while(true)} loop) allows the browser to conserve resources when the user is on a different tab by pausing the app entirely.
        It also takes advantage of knowing the display's refresh rate to avoid rendering faster than the display can actually update.
        Typical modern displays have 60 hz refresh rates corresponding to a maximum of 60 renders per second.
        The median frame rate over the 100 most recent renders can be displayed in the application if the \code{showFPS} setting is set to \code{true}. 
        
        The steps of the render loop are:
        \begin{enumerate}
            \item Check the UI state to determine if any values have changed (e.g., filter or colormap values, particle group on/off toggle, velocity vector checkbox, etc.) and update the particle meshes with new particle sizes/colors as needed.
            \item Draw the particles to either the canvas or to a buffer for remapping (projection mode) or saving to disk (take snapshot button).
            \item Check the loop duration and determine if the app should pause the loop if rendering is taking longer than 1.5 seconds/frame to avoid freezing the user's computer.
        \end{enumerate}
        
        \subsubsection{Render loop: checking the UI state and updating particle geometry and material objects}
            Because the application is interactive, users can change parameters of the visualization through the user interface and using the mouse and keyboard throughout the render loop.
            At the beginning of each render pass, changes to the UI are noted and and applied to the particle geometry and material objects as needed.
        
            In general, creating and disposing of \threejs/ geometry objects is (relatively) computationally expensive.
            So, instead of creating a new object whenever a UI element is changed we update all of the relevant geometry buffers registered to the \threejs/ scene in order to enact the corresponding change.
            Most changes do not impact performance.
            However operations that require looping over every particle in the Javascript interpreter, can potentially slow down the frame rate for large numbers of particles.
            
            The most frequent action that requires looping over every particle is changing the size and alpha values to show(/hide) a given particle when filter limits are applied.
            If the user moves the position of a filter handle, then the code loops through each particle to set its size and alpha values to zero if it is outside the filter range (or reset its size and alpha values to their original values if it was previously outside the filter boundaries but no longer is).
            As a result, moving the filter handles can temporarily introduce latency for large datasets (${\gtrsim} 10^6$ particles) until the filter limits are applied. 
            Changing the colormap variable (or toggling the colormap on/off) also requires looping over every particle in order to set the colormap scalar field value for each particle in the geometry buffer.
            After these expensive operations are completed, the framerate and interactivity return to normal.
    
        \subsubsection{Render loop: rendering the particles} \label{s:shaders}
            The \threejs/ point cloud objects are rendered to the canvas using WebGL vertex and fragment shaders.
            To do this there are, broadly speaking, three steps: 
            \begin{enumerate}
                \item determine each particle's position and size relative to the canvas 
                \item determine the color of each pixel the particle overlaps
                \item blend the colors according to a specified blending equation.
            \end{enumerate}
            
            The vertex shader interprets the data in order to position and scale the points in the camera's frame, accounting for camera distance, per-particle radial scaling, and the point size multiplier.
            It also applies a minimum and maximum point size to attempt to 1) ensure all particles are visible, even at great distance and 2) to decrease the likelihood that too many particles overlap and require blending.
            If the user enables velocity animation via the UI, the particle's position is offset in the direction of its velocity vector as defined by the input values in the UI (see Figure~\ref{f:particle_controls}f).
            
            The fragment shader defines the color for each pixel that the particle covers. 
            By discarding certain pixels we can draw different shapes at the location of each particle, e.g. circles or velocity vectors.
            Within the fragment shader, we can also: 1) apply a radial gradient in opacity (alpha) to give the particles a ``fuzzy'' edge or 2) change the color in a linear gradient along the velocity vector.
            
            If a colormap is enabled, the fragment shader will also read in the colormap texture (defined by a grid of colormaps stored in the \code{static/textures/colormaps.jpg}). 
            The colormap variable field value is remapped to the $[0,1]$ interval using the minimum and maximum colormap limits. 
            The remapped field value is used to sample the colormap texture in order to derive the appropriate color for the given particle.  
            So that RGBA values of overlapping particles do not blend and produce pixel colors that do not correspond to a value on the colormap, we change the blending function and enable the depth buffer such that each pixel takes the color of \textit{the particle nearest the camera} (i.e., enabling the depth buffer to perform a depth test and enforcing the ``normal'', rather than ``additive'', blending mode for that particle group).
                
            If column density projection mode is enabled, then particles are first rendered to a texture buffer rather than directly to the canvas. 
            The R channel of the texture buffer is used to accumulate the number of particles in each pixel using an additive blending function.
            The texture's R channel value is then remapped to an RGB color according to a colormap (selected from the drop-down menu) then rendered to the visible canvas.

        \subsubsection{Render loop: deciding if the app should sleep}
            If a rendering pass takes longer than 1.5 seconds, a sleep screen with a warning message is shown, and the app is paused until the user clicks again. 
            Depending on the size of the dataset (or more commonly, how many overlapping particles need to be blended on screen) the frame rate can drop.
            Because the \textit{entire browser} is unresponsive for the duration of a rendering pass, we enforce this test to prevent users from being locked out of closing the application (and to avoid crashing the users computer).
            The application can be woken and the render loop re-engaged by clicking on the sleep screen. 
                        
            The two most common causes of the sleep screen are: 1) switching browser tabs (in which case there is no issue) and 2) using a large point size multiplier (``sizeMult'') which causes many of the particles to overlap on top of one another and their colors to be blended (an expensive, but not valuable, calculation).
            In scenario 2), it is recommended that users either refresh their window to reset the point size multiplier to its default value, change the point size multiplier to a smaller value in the UI, or set a new initial size multiplier in the input settings \code{.JSON} file.

\section{Progressively loading large datasets with an octree} \label{s:octree}
    In order to visualize very large datasets (which can require more memory than is available to the browser to visualize in its entirety simultaneously), Firefly can load only that data which is nearby to and within view of the camera.
    To use this feature users must pre-format their data as an ``octree'' using the PDPP and write the  octree files to disk.
    An octree is a hierarchically structured scheme of partitioning space into incrementally smaller octants.
    Octrees are commonly used in cosmological simulations whose code makes use of adaptive mesh refinement (AMR) \citep[e.g.][]{Kravtsov1997,Teyssier2002} to represent astrophysical fluids on a grid rather than with (Lagrangian) particles.
    We borrow the concept and apply it to particle-based data in order to partition the dataset spatially into discrete (relatively) collocated chunks that can be identified and loaded on demand.
    
    Using a dataset that has been pre-formatted as an octree dramatically improves performance, reduces Firefly's start-up time to nearly instantaneous, reduces the browser's total memory usage, and allows the extent of the visualization domain to scale effectively infinitely (though importantly not all of the domain may be visualized at once). 
    Should a user decide to pre-format their data as an octree, a single function call in the PDPP wraps the construction of the tree (described below), produces the octree binary files, and registers them to the manifest file (described in Appendix~\ref{s:under_the_hood}). 
    
    We note that there are a few minor drawbacks to this approach.
    For one, the pre-formatting step can be computationally expensive for especially large datasets (see \S \ref{s:performance_tests} for a quantitative benchmark).
    Using octree formatted data also requires that the data be saved to disk (i.e., you cannot load octree data directly into Firefly even when run locally with a \flask/-enabled server). 
    Nonetheless, utilizing an octree can provide such a significant performance boost, and can enable the visualization of extremely large datasets that are otherwise inaccessible to interactive visualization techniques, that the benefits outweigh these minor drawbacks.  

    Inside the Firefly application, exploring the octree is (relatively) seamless for the end-user. 
    When a node is within the camera view and close enough that it overlaps multiple pixels it is ``opened'' and the raw particle data associated with that node is queued to be loaded. 
    On the other hand, when the node no longer overlaps multiple pixels or the camera pans such that it is offscreen, the raw particle data is purged from memory freeing it up to load new data.
    Nodes are also forcibly closed (oldest first) to enforce a fiducial memory limit of 2 GB to prevent the app from crashing (this value can be configured by users in the settings \code{.JSON} file).
    
    \subsection{Building the octree using the PDPP} \label{s:octree_algorithm}
        For our implementation of an octree, we typically group particles into nodes that contain $10^4-10^5$ particles each. 
        The minimum and maximum number of particles per node can be chosen by the user, and we explore the tree construction time for different values in \S \ref{s:performance_tests_octree}. 
        Throughout the remainder of this section, we refer to the ``raw'' particle data associated with a node as its ``buffer data.''
        Additionally, we also accumulate aggregate statistics while building the tree (like total mass and mass weighted scalar fields).
        Thus, along with pointers to any child nodes that may be associated with it, every node in the octree has its own \textit{aggregate data} computed over all the particle data below it that represents the node on average and its own \textit{buffer data} it is responsible for storing.
        
        The algorithm we employ begins by first measuring the extent of the 99\% of particles nearest to the center of the dataset to define the bounding box of the root node. 
        We hold the remaining furthest 1\% until after the octree has been constructed and it add as buffer data to the root node to minimize the effect of outliers on the structure of the tree (in particular, outliers can result in many empty layers of refinement that bloat the structure of the tree). 
        Then we sort all the particles into the eight sub-octants of the root node.
        Any sub-octant which contains more than the maximum allowable number of particles per node ``overflows'' and the buffer particles are sorted into the eight sub-octants of that node.
        Thus the octree is spatially refined as particles are iteratively sorted by recursively applying the maximum-threshold-to-overflow criteria for each of the resulting eight child nodes (and their children, etc.).
        At each refinement, we merge child nodes below the minimum required particles per node \textit{back into their parents} (a procedure we term ``pruning'').
        By pruning the octree we reduce the number of extraneous nodes which may contain only a handful of particles. 
        
        Because of the inherent asymmetry in coordinates of a typical dataset, it is frequently the case that some child nodes have a particle count below the minimum threshold while their siblings do not and may even have (grand-)children of their own.
        Thus, by pruning the octree it is possible, and in fact frequently the case, that a node has both buffer particle data it is responsible for and children. 
        
        The octree build implementation is optimized for extremely large datasets. 
        In particular, it can be run in parallel using multiple threads and does not require that the entire dataset be loaded into memory. 
        It does this by breaking up the dataset into ``work units'' which can be split across independent tasks and then synchronized after the refinement is complete. 
        These work units are distributed among the available worker threads using Python's \code{multiprocessing} module.
        
       The PDPP saves the octree's particle data in binary format, and also creates a \json/ file for Firefly that contains the octree structure, accumulated aggregate data for each node, and the locations of the node centers and centers of mass. 
        
    \subsection{Implementing an octree in Firefly}
        In the Firefly web application, an octree is represented as a Javascript object. 
        This object is defined by the octree \json/ file (from PDPP) and created in the ``Build Octree'' step of Figure \ref{f:startup_diagram}.
        An octree contains only a single particle group, though multiple octrees (one for each particle group that has an octree) can be loaded simultaneously. 
        Each node contains pointers to the files on disk, byte sizes, and byte offsets of its buffer data in addition to its center of mass, accumulated aggregate data, bounding box, and list of children. 
        Material and geometry mesh objects (described above in Appendix~\ref{s:under_the_hood}) are dynamically created and filled with this buffer data during the render loop when nodes are opened.
        
        The procedure of updating the octree in the render loop is visualized in the right half of Figure~\ref{f:combined_diagram}.
        We (un)load particle data asynchronously because the interactivity of the application is impacted by the time it takes to create (dispose) of \threejs/ geometry meshes.
        When there are multiple octrees (corresponding to different particle groups), Firefly will cycle through each particle group on consecutive render passes in order to (un)load data from each particle group with equal priority.
        To keep track of the buffer data that needs to be asynchronously loaded or purged, each octree has its own ``draw'' and ``remove'' queues.
        
        Each draw queue uses its own lock so that only one node's data is loaded at a time for a given particle group.
        At each render pass, if the draw queue is not empty and not locked, the node nearest to the camera in the draw queue is selected and the queue is locked.
        The particle data associated with that node is then loaded from the corresponding binary file and the node is removed from the draw queue. 
        Once the data is loaded from disk, new material and geometry objects are created, and the loaded data is copied into their buffers.
        Finally, the geometry mesh is added to the scene and the draw queue is unlocked.
        
        Because disposing of material and geometry objects is much less expensive than creating them and adding them to the scene, the entire remove queue is emptied at each render pass.
        To remove a node, the material and geometry objects are disposed of, and the Javascript array data is freed (this has the effect of releasing the memory back to the application which can be monitored by a memory usage counter that can be toggled on and off with a flag in the input settings \json/ ). 
        
        After dealing with the draw and remove queues, we then update each of the queues by walking the current particle group's octree.
        At each node, we decide whether that node should be opened, closed, or ignored.
        If the current node is on screen and covers more than one pixel from the camera perspective, the node is added to the particle group's draw queue.
        If the current node has already been drawn but subtends less than 1 pixel on screen, the node is added to the particle group's remove queue.
        If a node is added to either queue but also exists in its opposing queue then it is removed from the \textit{older} queue (for example if a node was on screen and added to the draw queue but then exited the screen and no longer needs to be drawn).
        
        Through this method, Firefly is able to render and \textit{interactively} explore exceptionally large data sets.  
        We provide one example of interacting with nearly 1.5$\times10^9$ stars from Gaia DR3 in \S\ref{s:gaia_example} in real time, which can only be achieved using an octree.  
        As data sets continue to grow, Firefly's octree rendering method will be a highly valuable tool for data exploration and public outreach. 
 
\end{document}